\documentclass[letterpaper,twocolumn,10pt]{article}
\usepackage{hyperref}
\usepackage[hyphenbreaks]{breakurl}
\usepackage{amssymb}
\usepackage{usenix2019_v3}
\usepackage{filecontents}
\usepackage{balance}
\usepackage{amsmath}
\usepackage{url}
\usepackage{color}
\usepackage{caption}
\usepackage{subcaption}
\usepackage{multirow}
\usepackage{booktabs}
\usepackage{tikz}
\usepackage{epsfig,endnotes}
\usepackage{times}
\usepackage{algorithm,algpseudocode}
\usepackage{tablefootnote}
\usepackage{threeparttable}
\usepackage{booktabs}
\usepackage{longtable}
\usepackage{enumitem}
\usepackage{float}

\newtheorem{defi}{Definition}

\algnewcommand\algorithmicforeach{\textbf{for each:}}
\algnewcommand\ForEach{\item[ \algorithmicforeach]}
\algnewcommand\algorithmicendforeach{\textbf{end for}}
\algnewcommand\EndForEach{\item[ \algorithmicendforeach]}
\definecolor{ocean}{RGB}{78, 168, 245}
\newcommand{\bit}[1]{\textit{\textbf{#1}}}
\newcommand{\para}[1]{{\vspace{2pt} \noindent \textbf{#1}
    \hspace{6pt}}}

\newcommand{\spara}[1]{{\vspace{1pt} {\em #1}
    \hspace{3pt}}}

\newcommand{\todo}[1]{{\color{red} \textbf{TODO:} #1}}
 \newcommand{\emily}[1]{{\color{black} #1}}

\newcommand{\htedit}[1]{{\color{black} #1}}

\newcommand{\revision}[1]{{\color{black} #1}}

\newcommand{\secspace}{\vspace{-0.1in}}

\newcommand{\adv}{Adv}

\newcommand{\mnist}{{\tt MNIST}}
\newcommand{\cifar}{{\tt CIFAR-10}}
\newcommand{\speech}{{\tt SSCD}}
\newcommand{\har}{{\tt WISDM}}
\newcommand{\gtsrb}{{\tt GTSRB}}

\newcommand{\youtubeface}{{\tt YTFaces}}

\newcommand{\etal}{{\em et al.\ }}

\newenvironment{packed_itemize}{
\begin{list}{\labelitemi}{\leftmargin=1em}
  \setlength{\itemsep}{1pt}
  \setlength{\parskip}{0pt}
  \setlength{\parsep}{0pt}
  \setlength{\headsep}{0pt}
  \setlength{\topskip}{0pt}
  \setlength{\topmargin}{0pt}
  \setlength{\topsep}{0pt}
  \setlength{\partopsep}{0pt}
}{\end{list}}

\newcommand{\model}{{\mathbf{F}}}

\newcommand{\el}{{\ell_{\model}}}

\newfont{\mycrnotice}{ptmr8t at 7pt}
\newfont{\myconfname}{ptmri8t at 7pt}

\begin{document}


\title{\vspace{-3cm}Piracy Resistant Watermarks for Deep Neural Networks}
\author{Huiying Li, Emily Wenger, Shawn Shan, Ben Y. Zhao, Haitao
  Zheng\\
  {\em Department of Computer Science, University of Chicago} \\
  {\em \{huiyingli, ewillson, shansixiong, ravenben, htzheng\}@cs.uchicago.edu}}

\maketitle

\begin{abstract}
  As companies continue to invest heavily in larger, more accurate and more
  robust deep learning models, they are exploring approaches to monetize
  their models while protecting their intellectual property.
  Model licensing is promising, but requires a robust tool for owners to claim
  ownership of models, {\em i.e.} a watermark. Unfortunately, current
  designs have not been able to address {\em piracy attacks}, where third
  parties falsely claim model ownership by embedding their own ``pirate
  watermarks'' into an already-watermarked model.

  We observe that resistance to piracy attacks is fundamentally at odds with
  the current use of incremental training to embed watermarks into models.
  In this work, we propose {\em null embedding}, a new way to build
  piracy-resistant watermarks into DNNs that can only take place at a model's
  initial training. A null embedding takes a bit string (watermark value) as
  input, and builds strong dependencies between the model's normal
  classification accuracy and the watermark.  As a result, attackers cannot
  remove an embedded watermark via tuning or incremental training, and cannot
  add new pirate watermarks to already watermarked models.  We
  empirically show that our proposed watermarks achieve piracy resistance and
  other watermark properties, over a wide range of tasks and
  models. Finally, we explore a number of adaptive counter-measures, and show
  our watermark remains robust against a variety of model modifications,
  including model fine-tuning, compression, and existing methods to
  detect/remove backdoors. Our watermarked models are also amenable to
  transfer learning without losing its watermark properties.
 \end{abstract}

\secspace
\section{Introduction}
\label{sec:intro}
\vspace{-0.05in}

State-of-the-art deep neural networks (DNNs) today are incredibly expensive
to train. For example, a new conversational model from Google Brain includes
2.6 billion parameters, and takes 30 days to train on 2048 TPU
cores~\cite{chatbot}. Even ``smaller'' models like ImageNet require
significant training (128 GPUs for 52 hours) to add robustness properties.

As training costs continue to grow with each generation of models, providers
must explore approaches to monetize models and recoup their training costs,
either through Machine Learning as a Service (MLaaS) platforms ({\em
  e.g.\/}~\cite{ribeiro2015mlaas, yao2017complexity}) that host models, or
fee-based licensing of pretrained models.  Both have serious
limitations. Hosted models are vulnerable to a number of model inversion or
inference attacks ({\em e.g.\/}~\cite{fredrikson2015model,
  tramer2016stealing, wang2018stealing}), while model licensing requires a
robust and persistent proof of model ownership.

DNN
watermarks~\cite{chen2019deepmarks,uchida2017embedding,zhang2018protecting}
are designed to address the need for proof of model ownership.  A robust
watermark should provide a persistent and unforgeable link between the model
and its owner or trainer. Such a watermark would require three properties. {\em First},
it needs to provide a strongly verifiable link between an owner and the
watermark ({\em authentication}). {\em Second}, a watermark needs to be
persistent, so that it cannot be corrupted, removed or manipulated by an
attacker ({\em persistence}). {\em Finally}, it should be unforgeable, such
that an attacker cannot add additional watermarks of their own to a model in
order to dispute ownership ({\em piracy-resistance}).


Despite a variety of approaches, current proposals have failed to achieve the
critical property of piracy resistance.  Without it, a user of the model can
train their own ``valid'' watermark into an already watermarked model,
effectively claiming ownership while preserving the model's classification
accuracy. \htedit{Specifically},  recent work~\cite{wang2019attacks} showed that regularizer-based
watermarking methods~\cite{uchida2017embedding, chen2019deepmarks,
  darvish2019deepsigns} were all vulnerable to piracy attacks. More
recent \htedit{watermark designs} rely on embedding classification artifacts into
models~\cite{zhang2018protecting, adi2018turning}. Unfortunately, our own
experiments show that both techniques can be overcome by
\htedit{successfully embedding} pirate watermarks
with moderate training.

But what makes piracy resistance so difficult to achieve? The answer is that
neural networks are designed to accept incremental training and
fine-tuning. DNNs can be fine-tuned with existing training data, trained to
learn or unlearn specific classification patterns, or ``retargeted'' to
classify \htedit{input} to new labels via transfer learning. In fact,
\htedit{existing designs of DNN} watermarks
rely on this incremental training property to embed themselves into models.
Thus, it is unsurprising that with additional effort, an attacker can use the
same mechanism to embed more watermarks into an already watermarked model.


In this work, we \htedit{propose} {\em null embedding}, a new approach for
embedding piracy-resistant watermarks into deep neural networks. Null
embedding does not rely on incremental training. Instead, it can only be
trained into a model at time of initial model
training. \htedit{Formally speaking}, a
null embedding \htedit{(parameterized by a bit
string)} imposes an additional constraint on the optimization process
used to train a model's normal classification behavior, \htedit{{\em i.e.\/} the classification rules used to
classify normal input. As this constraint is imposed at time of initial
model training, it builds strong dependencies between
normal classification accuracy and the given null embedding parameter.
After a model is trained with a given null embedding, further
(incremental) training to add a new null embedding fails,
because it generates conflicts with the existing null embedding and} destroys the model's normal classification accuracy.



Based on the new null-embedding technique, we propose a strong watermark
system that integrates public key cryptography and verifiable signatures into
a bit-string embedded as a watermark in a DNN model. The embedded bit string
inside a watermarked model is easily identified and \htedit{securely} associated with the
model owner. \htedit{The presence of the watermark does not affect the
  model's normal classification accuracy.}  More importantly, attempts to train a different, pirate watermark into
a watermarked model would destroy the model's value, {\em i.e.\/} its
ability to classify normal inputs. This deters any piracy
attacks against watermarked models.


Our exploration of the null-embedding watermark produces
several key findings, which we summarize below:
\begin{packed_itemize} \vspace{-0.07in}
\revision{
\item We define the property of piracy resistance for DNN watermarks, and
  empirically confirm that existing approaches are vulnerable to piracy attacks.
\item We validate the null-embedding technique and associated watermark on
  six classification tasks and different model architectures, extending
  beyond image classification to speech and human activity recognition.} We
  show that piracy attacks against our watermarked models actually destroy
  model classification properties, and are no better than training the model
  from scratch, regardless of computation effort (\S\ref{subsec:null},
  \S\ref{subsec:evalpirate}).  We also confirm that our watermarks 
  achieve all basic watermark properties (\S\ref{subsec:evalbasic}).
\item We show that watermarked models are amenable to transfer
  learning systems: models can learn classification of new labels without losing
  their watermark properties (\S\ref{sec:transfer}). 
\item With respect to countermeasures, we show our watermarks cannot be
  removed by modifications such as model fine-tuning, neuron pruning, model
  compression, or backdoor detection methods (\S\ref{subsec:common},
  \S\ref{subsec:backdoor_detection}). They disrupt the model's normal
  classification before they begin to have any impact on the watermark.
  \revision{We discuss model extraction attacks, and explore their limited
    benefits and practical costs using empirical results (\S\ref{subsec:stealing}).}
\vspace{-0.06in}
\end{packed_itemize}
Overall, our empirical results show that null-embedding shows promise as a way to
embed watermarks that resist piracy attacks.  We discuss
limitations and future work in \S\ref{sec:discussion}. 




\vspace{-0.15in}
\section{Related Work}
\vspace{-0.06in}
\label{sec:back}
The goal of watermarking is to add an unobtrusive and tamper-resistant
signal to the host data, such that it can be reliably recovered from the host data using a
recovery key.  As background,  we now summarize existing works on digital watermarks,
which have been well studied for multimedia data and recently
explored for deep neural networks.

\secspace
\subsection{Digital Watermarks for Multimedia Data}
Watermarking multimedia data has been widely
studied in the literature ({\em e.g.\/} a
survey~\cite{hartung1999multimedia}). A watermark can be added to {\em
  images} by embedding a low-amplitude, pseudorandom signal on
top of the host image. To minimize the impact on the host, one can
add it to the least significant bits of grayscale
images~\cite{van1994digital}, or leverage various types of statistical
distributions and transformations of the image ({\em
  e.g.\/}\cite{tanaka1990embedding, kutter1997digital,
  bender1996techniques}).  For video, a watermark can take the
form of imperceptible perturbations of wavelet coefficients for each
video
frame~\cite{swanson1998multiresolution} or employ other perception
measures to make it invisible to
humans~\cite{wolfgang1999perceptual}. Finally, watermarks can be
injected into audio by modifying its Fourier coefficients~\cite{bender1996techniques,tilki1996encoding,swanson1998robust}.

\secspace
\subsection{Digital Watermarks for DNNs}

Recent works have examined the feasibility of injecting watermarks into DNN models. They can be
divided into two groups based on the embedding methodology.


\para{Weights-based Watermarks.} The first
group~\cite{uchida2017embedding, chen2019deepmarks,
  darvish2019deepsigns} embeds
watermarks directly onto model weights, by adding a
regularizer containing a statistical bias during training.
But anyone knowing
the methodology can extract and remove the injected
watermark without knowing the secret used to inject it.  For example, a recent attack shows that these watermarks
can be detected and removed by overwriting the statistical
bias~\cite{wang2019attacks}. Another design~\cite{fan2019rethinking}
enables ``ownership verification'' by adding
special ``passport'' layers into the model, such that the model performs poorly when
passport layer weights are not present. \htedit{This design relies on the secrecy of
  passport layer weights to prove model ownership.  Yet the paper's own
  results show attackers can reverse engineer a set of effective  passport layer
weights.  Since there is no secure link between these weights and the owner,
attackers can reverse engineer a set of valid weights and claim ownership.}


\para{Query-based Watermarks.} \revision{DAWN~\cite{szyller2019dawn}
  proposes to protect an online, proprietary model by selectively
  mislabeling query responses.  If 
  an attacker attempts a model extraction attack against the hosted model,
  these modified responses leave a detectable ``mark'' inside the new
  model. This is used as a ``watermark'' to prove that the model was stolen.
}

\para{Classification-based Watermarks.} The second approach embeds
watermarks in model classification results.  Recent
work~\cite{zhang2018protecting} injects watermarks using the backdoor attack method, where applying a specific ``trigger''
pattern (defined by the watermark) to any input will
produce a model misclassification to a specific target label.
However, backdoor-based watermarks can be removed using existing
backdoor defenses ({\em e.g.\/}\cite{wang2019neural}), even without knowing the trigger.
Furthermore, this proposal provides no verifiable link between
the trigger and the identity of the model owner. Any party who
discovers the backdoor trigger in the model can claim they inserted
it, resulting in a dispute of ownership.

Another work~\cite{adi2018turning} uses a slightly different approach. It
trains watermarks as a set of classification rules associated with a
set of self-engineered, abstract images only known to the
model owner. Before embedding this (secret) set of images/labels into the model,
the  owner creates a set of commitments over the image/label
pairs. By selectively revealing these commitments and showing that
they are present in the model, the owner proves
their ownership.


\secspace
\section{Problem Context and Threat Model}
\vspace{-0.06in}
\label{subsec:threat}
To provide context for our later discussion, we now describe the
problem setting and our threat model.

\para{Ownership Watermark.}
Our goal is to design a robust {\em ownership watermark}, which
proves with high probability that a specific watermarked DNN model was
created by a particular owner O. Consider the following scenario. O
plans to train a DNN model $\mathbf{F}_\theta$ for a specific task,
leveraging significant resources to do so (e.g. training data and
computational hardware). O wishes to license or otherwise share this valuable model with others, either directly
or through transfer learning, while maintaining
ownership over the intellectual property that is the model. If ownership of
the model ever comes into question, O must prove
that they and only they could have created $\mathbf{F}_\theta$. To
prove \emily{their} ownership of $\mathbf{F}_\theta$ on demand, O embeds watermark $\mathbb{W}$ into
the model simultaneously when training the model. This watermark needs to be robust against attacks by
a malicious adversary $\adv$.


%

%

\para{Threat Model.} At a high level, the adversary $\adv$ wants to
stake its own ownership claims on $\mathbf{F}_\theta$ or at least
destroy O's claims.  We summarize possible adversary goals as follows:
\begin{packed_itemize} \vspace{-0.05in}
\item {\bf Corruption}: $\adv$ corrupts or removes the watermark $\mathbb{W}$,
  making it unrecognizable and removing O's ownership claim.
\item {\bf Piracy}: $\adv$ adds its own watermark $\mathbb{W}_A$ so it
  can assert its ownership claim alongside O's.
 \item {\bf Takeover}: A stronger version of piracy is that $\adv$ replaces $\mathbb{W}$ with its own watermark
  $\mathbb{W}_A$, in order to completely take over ownership claims of
  the model.
  \vspace{-0.05in}
\end{packed_itemize}

\revision{We make three assumptions about the adversary.  {\em First}, we assume the
owner makes the watermark procedure public and known to all, and $\adv$ will
follow the same procedure in trying to claim ownership of the model. {\em
  Second}, $\adv$ is not willing to sacrifice model functionality, {\em
  i.e.\/} the attack fails if it dramatically lowers the model's normal
classification accuracy.
{\em Third},} $\adv$ has limited training data and finite
computational resources. If $\adv$ has as much or even more training data
as O, then it would be easier to train its own model from scratch, making
ownership questions over $\mathbf{F}_\theta$ irrelevant. We assume finite resources, because
at some point, trying to compromise the watermark will be more costly
in terms of computational resources and time than training a model from scratch. Our goal is to make compromising a
watermark sufficiently difficult, such that it is more cost-efficient for an
adversary to pay reasonable licensing costs instead.

\revision{
\begin{table*}[t]
  \centering
  \resizebox{0.98\textwidth}{!}{
    \begin{tabular}{|l|c|c|c|c|c|c|}
      \hline
      \multirow{2}{*}{{\bf Task}} & \multicolumn{3}{c|}{ {\bf Watermark
                              Design~\cite{adi2018turning}}}
           &      \multicolumn{3}{c|}{ {\bf Watermark Design~\cite{zhang2018protecting}}} \\ \cline{2-7}
      & {\bf Normal Classification}      & {\bf  Original Watermark (\%) }&
                                                                 {\bf Pirate
                                                                 Watermark
                                                                 (\%)}
                                    &     {\bf  Normal   Classification
                                      (\%) } & {\bf Original Watermark (\%)}
                                    & {\bf Pirate Watermark (\%) }\\ \hline
      \mnist  & $98.5 \pm 0.0$ / $97.4 \pm 0.0$    & $
                                                                 100.0
                                                                 \pm
                                                                   0.0$/$
                                                                   48.8

                                                                   \pm
                                                                   0.4$
      & \textbf{$93.0 \pm 0.0$} & $98.6 \pm 0.1$ /
                                         $98.5 \pm  0.1$ & $100.0 \pm
                                                     0.0$ / $100.0
                                                     \pm 0.0$  &
                                                                   \textbf{$100.0
                                                                    \pm
                                                                    0.0$}  \\ \hline
      \youtubeface    & $ 98.4 \pm 0.5$ / $95.4 \pm 0.8$
           & $99.8 \pm 0.7$/ $53.3 \pm 6.1$ &
                                                             \textbf{$98.0
                                                             \pm
                                                        0.0$} &
                                                                  $98.4
                                                                  \pm 0.2$
                                                                  /
                                                                  $98.1
                                                                  \pm 0.2$
                                                               &
                                                                 $100.0
                                                                 \pm 0.0$
                                                                 /
                                                                 $83.2
                                                                 \pm 34.4$
                                    & \textbf{$100.0 \pm 0.0$}  \\ \hline
      \gtsrb  & $96.0 \pm 0.1 $ / $95.7 \pm 0.1$ &
                                                            $100.0
                                                             \pm 0.0$
                                                                /
                                                                $98.0
                                                                \pm
                                                                0.0$
      & \textbf{$98.0 \pm 0.0$} & $96.1 \pm 0.1$ / $
                                            96.0 \pm 0.2$ &
                                                                  $100.0
                                                                  \pm
                                                                  0.0
                                                                  $ / $
                                                                  100.0
                                                                  \pm
                                                                  0.0$         &
                                                                               \textbf{$100.0 \pm 0.0$}  \\ \hline
      \cifar   & $84.7 \pm 0.2$ / $ 84.0 \pm 0.2$ &
                                                                   $100.0 \pm 0.00$ / $98.2 \pm 1.47$
      & \textbf{$98.0 \pm 0.0$} & $86.0 \pm 0.4 $ / $ 85.6 \pm 0.3$ & $100.0\pm0.0$ / $100.0 \pm  0.0$ & \textbf{$100.0 \pm 0.0$}  \\ \hline
    \end{tabular}
   }
    \vspace{-.1in}
    \caption{\revision{{\em  Performance of watermarked models for four classification
          tasks. We show the before / after a piracy attack results in
          the table.  
          The metrics are model classification accuracy on normal inputs,
      classification accuracy of the original (or owner) watermark, and
      classification accuracy of the pirate
      watermark.  These results show that, for both watermark
      designs, an attacker can
      successfully insert a new, verifiable pirate watermark on a watermarked
      model. }}
    }
    \label{table:prework_piracy_std}
    \vspace{-.1in}
\end{table*}
}

\secspace
\section{Understanding Piracy Resistance}
\vspace{-0.05in}
\label{sec:piracy}
\revision{While {\em piracy resistance} is  a
critical requirement for DNN watermarks, it is not addressed in prior work. To
the best of our knowledge, all existing works are
vulnerable to piracy attacks.} \emily{In this section, we empirically
validate this vulnerability,} discuss why existing designs
fail to achieve piracy resistance, and propose \emily{an alternative design}.

\secspace
\subsection{The Need for Piracy Resistance}
In an {\em ownership piracy} attack, an attacker attempts to embed their watermark into
a model that is already watermarked. If the attacker can
successfully embed their watermark into the watermarked model, the
owner's watermark can no longer prove their (unique) ownership. That is,
the ambiguity introduced by the presence of multiple watermarks
invalidates the true owner's claim of ownership. To be effective, a
DNN watermark {\em must} resist ownership piracy attacks.

\secspace
\subsection{Existing Works are Not Piracy Resistant}
We show that, unfortunately, all existing DNN watermarking schemes
are vulnerable to ownership piracy attacks.


\para{Piracy Resistance of Weights-based Watermarks.} Recent
work~\cite{wang2019attacks} already proves that
regularizer-based watermarking methods~\cite{uchida2017embedding, chen2019deepmarks, darvish2019deepsigns} are vulnerable to ownership
piracy attacks, {\em i.e.\/} an attacker can inject new watermarks
into a watermarked model without compromising the model's normal
classification performance.  Furthermore, the injection of a new
watermark will largely degrade or even remove the original
watermark.
Another watermark design in this category~\cite{fan2019rethinking} also fails to achieve
piracy resistance because it cannot securely link an embedded
watermark to the model owner. An attacker can demonstrate the existence
of a pirate watermark without embedding it into the model.

\revision{
  \para{Piracy Resistance of Query-based Watermarks.}
  While DAWN~\cite{szyller2019dawn} could ``mark''  models (re)created using a model extraction attack, it cannot prevent adversaries
  from inserting an {\em additional} mark to dispute model
  ownership.
  DAWN states that if more than one marks are found in a
  model, ownership will default to the mark that was registered
  first. However, if the attacker registers their mark before the true owner registers theirs, the
  attacker succeeds in pirating the model.
}



\para{Piracy Resistance of Classification-based Watermarks.}
In the
following section, we show empirically that existing
works~\cite{zhang2018protecting, adi2018turning} are vulnerable to
piracy attacks.  We follow the original papers~\cite{zhang2018protecting, adi2018turning}  to re-implement the
proposed watermarking schemes on four classification tasks (\mnist,
\youtubeface, \gtsrb, and \cifar)\footnote{\revision{Both~\cite{zhang2018protecting, adi2018turning} design and evaluate watermarks for
image-based classification tasks including \cifar{} and \mnist.}}. Details of these
tasks are listed in \S\ref{sec:expr_setup}.
Additional details concerning the DNN model architectures,
training parameters, and watermark triggers used in our experiments
can be found in the Appendix~\ref{sec:exp_config_pre}.

To implement piracy attacks, we assume a strong attacker who has access
to 5,000 original training images and the
watermarked model. The goal of the attacker is to inject
a new, verifiable pirate watermark into the model.  This is achieved
by the attacker updating the model using training data related to
the pirate watermark. We found that for all four DNN models,
  a small number of training epochs is sufficient to successfully
  embed the pirate watermark. For both~\cite{zhang2018protecting}
  and~\cite{adi2018turning}, \mnist, \gtsrb, and \cifar~only need
  10 epochs while \youtubeface~only needs 1 epoch.






To evaluate each method's piracy resistance, we use three metrics: (1) the model's normal
classification accuracy,  (2) its  classification accuracy on the
original (owner) watermark, and (3) its classification accuracy on the pirate
watermark. We record these before and after the piracy attack to
measure the impact of the attack. \htedit{In an ideal watermark design, no
  piracy attack should be able to successfully embed a pirate watermark into
  a model while maintaining its classification accuracy for normal inputs.}


We list the results in Table~\ref{table:prework_piracy_std}.
For both watermark designs, piracy attacks succeed (are recognized
consistently) across all four classification tasks, and introduce minimal
changes to the normal classification accuracy. For some models,
the piracy attack also heavily degrades
the original watermark. These results show that
existing watermark designs are vulnerable to piracy attacks.

\para{Note on the Piracy Claim in~\cite{adi2018turning}.}
\cite{adi2018turning} assumes that the adversary uses the same number of
watermark training epochs as the model owner, and applies {\em an additional
verification step} via fine tuning, and claims the original watermark is more robust
against fine-tuning than the pirate watermark.
\emily{For completeness, we perform additional tests that reproduce the exact
  experimental configuration (same number of original/pirate watermark
  training epochs, followed by 10 epochs of fine-tuning) as~\cite{adi2018turning}.}
Contrary to~\cite{adi2018turning}, our results
show that across all four tasks, the pirate watermark is {\em more}
robust to fine-tuning than the original watermark. For \mnist,
\cifar, and \gtsrb, \revision{the pirate watermark's classification
accuracy remains on average $82\% \pm 3\%$ after
fine-tuning, while the accuracy of the original watermark drops to
$57\% \pm 2\%$}.


\secspace
\subsection{Rethinking Piracy Resistance}

The key obstacle to piracy resistance is the \htedit{{\em incremental
    trainability}} property inherent to DNN models. A pretrained
model's parameters can be further tweaked by
fine-tuning the model with more training data. Such fine-tuning can be
designed to not disturb the foundational classification rules  used to accurately classify normal inputs, but change
fine-grained model behaviors beyond normal classification, {\em
  e.g.\/} adding new classification rules
related to a backdoor trigger.

\para{Existing Watermark Methodology:
Separating Watermark from Normal
  Classification.} Existing watermark designs, particularly classification-based
watermarks,  leverage the {\em incremental trainability}
property to inject watermarks. In these designs, the model's normal
classification behaviors are made {\em independent} of the
watermark-specific behaviors.  Thus, the foundational
classification rules learned by the model to classify normal inputs
will not be affected by the embedded watermark. Such independence or
isolation allows an adversary to successfully embed new (pirate) watermarks into
the model without affecting normal classification.

\htedit{
\para{Our New Methodology: Using Watermark to Control Normal
  Classification.}  Instead of separating watermark
from  normal classification,  we propose to use the ownership
watermark  to constrain (or regulate) the generation/optimization of normal classification
rules. Furthermore, this
constraint is imposed at time of initial model training, creating
strong dependencies between normal classification accuracy and the
specific bit string in the
given watermark.  Once a model is trained / watermarked,  further
(incremental) training to add a new (pirate) watermark will break the
model's normal classification rules. Now the updated model is no
longer useful, making the
piracy attack irrelevant.

A stubborn adversary can continue to apply more training to
``relearn'' normal classification rules under the new constraint
imposed by the pirate watermark.  Yet the corresponding training cost
is significantly higher ({\em e.g.\/} by a factor of 10 in our experiments)  than training the model (and adding the pirate
watermark) from scratch.   Such significant (and unnecessary) cost leaves no incentive
for piracy attacks in practice.

}

\secspace
\section{Piracy Resistance via Null Embedding}
\label{sec:intuition}

Following our new methodology, we now describe
``null embedding'',  an effective method to implement watermark-based
control on the generation of normal classification rules. In a
nutshell, null embedding adds a global, watermark-specific optimization
constraint on the search for normal classification rules. This
effectively projects the optimization space for normal classification
to a watermark-specific area.  Since different watermarks create
different constraints and thus projections of optimization space,
embedding a pirate watermark to a watermarked model will create
conflicts and break the model's normal classification.

In this section, we describe the operation of
null embedding and its key properties that allow our watermark
  to achieve piracy resistance. \htedit{Then,
we show that a practical DNN watermark needs to be ``dual embedded''
(via both true and null embedding)  into
the model to achieve piracy resistance
and be effectively verified and linked to the owner.}

\revision{Without loss of generality,  we describe both null embedding and our
  watermark design for
  CNN-based image classification tasks. Here the input to
  the model is an image. Our design can potentially be generalized to other
  classification tasks where the input to the classification model is a matrix.  Later in
  \S\ref{sec:eval} we implement and evaluate six classification tasks
  including image classification, speech recognition,  and human activity
  recognition via accelerometer data, showing that our proposed watermark design
can be successfully applied to all of them.}



\secspace
\subsection{Null Embedding}
\label{subsec:null}
Given a watermark sequence ({\em e.g.\/} a 0/1 bit string), the
process of null embedding \emily{uses this}  sequence to modify the effective optimization space used to train the model's
normal classification rules.  This is achieved by imposing a
constraint during training,  preventing a specific
\htedit{configuration on model inputs}
from affecting the normal  classification outcome.    To do so, null
embedding must take place at the time of original model training. The
model owner will start \emily{with an untrained} model, generate extra training data related to null
embedding, and train the model using the original and extra training
data.

\begin{figure}[t]
\centering
  \includegraphics[width=0.4\textwidth]{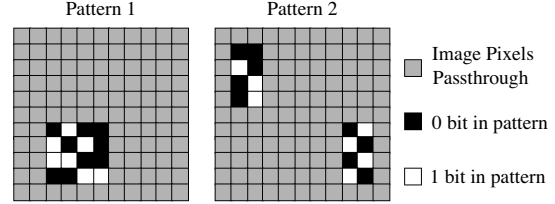}
  \vspace{-0.05in}
  \caption{\em Two examples of a null embedding pattern. For each filter
    pattern, the color of a pixel represents the value of that pixel in the filter: gray means
    no change (value -1), black means $0$ and white means $1$.   Each
    filter pattern is defined by the spatial distribution of the
    black/white pixel areas and the bit pattern in each black/white
    area.
  }
  \label{fig:filter}
\end{figure}

We formally define the process as follows.  Let $\mathbf{F}_\theta: \mathbb{R}^N \to \mathbb{R}^M$ be a DNN model that
maps an input $x\in\mathbb{R}^N$ to an output $y\in\mathbb{R}^M$.
Let a watermark pattern $p$ be a filter pattern applied
to an image. Two samples are shown in
Figure~\ref{fig:filter}.   Each filter pattern is defined by the
placements of the black (0) and white (1) pixels on top of the gray
background pixels (-1).


\vspace{-0.05in}
\begin{defi}[Null Embedding]
  \label{def:nullembedding} Let $\lambda$ be a
  very large positive value ($\lambda \rightarrow \infty$).  A
  filter pattern $p$ is successfully null-embedded into a DNN model
 $\mathbf{F}_\theta$ iff
  \begin{equation}\label{eq:nulldef}
    \mathbf{F}_{\theta} (x\oplus [p,\lambda]) =
    \mathbf{F}_{\theta}(x)=y, \quad \forall x\in \mathbb{R}^N,
  \end{equation}
where $y$ is the true label of $x$.  Here $x \oplus [p,\lambda]$ is an input filter operation.
  For each white (1)
  pixel of $p$,  it
  replaces the pixel of $x$ at the same position with $\lambda$; for
  each black (0) pixel of $p$, it replaces the corresponding pixel of
  $x$ with $-\lambda$;   the rest of $x$'s pixels
  remain unchanged.
\end{defi}\vspace{-0.05in}

This shows that when $p$ is successfully null embedded into the
model, \htedit{changing a set of $p$-defined pixels on {\em any} input $x$ to hold extreme values
$\lambda$ and $-\lambda$  would not change the classification
outcome.  This condition (and the use of extreme values) set a strong
and deterministic constraint on the
optimization process used to  learn the normal classification
rules. And by enforcing the constraint defined by (\ref{eq:nulldef}),  null embedding of a pattern $p$ will  {\em
  project} the model's effective input-vs-loss space ({\em i.e.\/} the optimization
landscape) into a sub-area defined by $p$.
}



\para{Properties of Null Embedding.}  We show that null embedding
displays two properties that help us design pirate resistant
watermarks. We also verify these properties empirically.

\vspace{3pt}
\bit{Observation 1}: {\em When $N_{p}$, the number of white/black pixels in $p$, is reasonably
  small,  null embedding $p$ into a model does not affect the model's
  normal classification accuracy.}
\vspace{3pt}

Null embedding of $p$ confines the model's optimization
landscape into a sub-area.  As long as this sub-area is sufficiently
large and diverse,  one can train the model to reach the desired
normal classification accuracy.  Our hypothesis is that when $N_p$ is
reasonably small compared to the size of the input image,  the
sub-area defined by $p$ would be sufficiently large and diverse to learn accurate
normal classification.


\revision{We test this hypothesis on the same \revision{four} image
classification tasks used in \S\ref{sec:piracy}  (\mnist, \youtubeface, \gtsrb, and
\cifar) and two non-image classification tasks (\speech~and \har).}  We
find that for
all of them, $N_p$ can be as large as 10\% of
the total input size without causing noticeable impact on normal classification accuracy.  For example,
$N_P=6\times6$ on $28\times 28$ images results in only 0.1\%-1.5\% accuracy loss.  One can potentially reduce this loss
by optimizing the design of filter pattern, {\em e.g.\/} configuring
white/black pixel area as irregular shapes, which we leave to future
work.


\vspace{5pt}
\bit{Observation 2}:  {\em Once a model is trained and null embedded
  with $p$,  an adversary cannot null embed a pirate $p'$ ($p'\neq p$)
  without largely degrading the model's normal classification accuracy.}
\vspace{5pt}

Our hypothesis is that null embedding of different patterns will create
different projections of the optimization space. Once a model is
successfully trained on $p$-based optimization space,  any attempt to
move
it to a different optimization space (defined by $p'$) will
immediately break the
model.

\begin{figure}[t]
  \centering
  \includegraphics[width=0.4\textwidth]{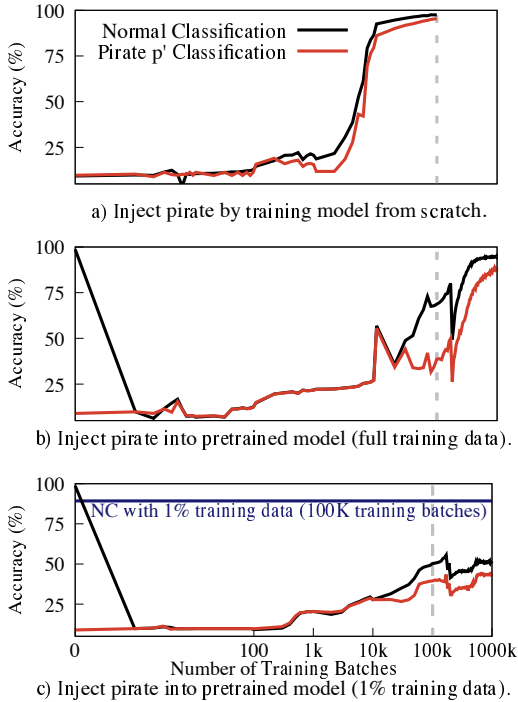}
  \vspace{-0.0in}
  \caption{\em \revision{An instance of \mnist{} demonstrating the
      significant cost of piracy attack on null embedding. We plot the
      normal classification accuracy and pirate watermark $p'$'s
      classification accuracy under three scenarios:  (a) injecting
      $p'$ by training the model from scratch;  (b) injecting $p'$ into an
      watermarked-model using fine-tuning and  full training
      data;  and (c)  same as (b) but using 1\% of training
      data.}}
  \label{fig:acc_change_percent}
  \vspace{-0.1in}
\end{figure}


\revision{We verify this hypothesis by studying a watermarked
  model's normal classification accuracy and pirate $p'$ classification accuracy as the
adversary fine-tunes the model to embed $p'$.  Our key finding is that
embedding $p'$ to a watermarked-model requires
significantly more computation than training a model from scratch
to embed $p'$.}

\revision{An example trace on \mnist{} is shown in
  Figure~\ref{fig:acc_change_percent}, where we compare three
  scenarios to add a watermark $p'$ to a model.  These are: (a) the
  adversary trains the model {\em from scratch} and
  injects $p'$,  (b) the adversary finds a watermarked model,
  fine-tunes the model to inject $p'$, using all the training
  data, and (c) which is the same as (b) except it uses 1\% of the
  training data.   Here we plot the training cost as the number of
  training batches (rather than epochs) to illustrate a fine-trained model
  behavior.  Under scenario (a), the training takes 100k batches to reach
  the proper accuracy levels.  Under scenarios (b) and (c), the first
  few batches of fine-tuning immediately drop the normal classification
  accuracy to 10\% (random guess). Even after 2k
  batches of fine-tuning,  both normal and $p'$'s classification
  accuracies are still low.   To reach the same accuracy as (a), the
  adversary will need 10x more training cost even when having the full
  training data (see Figure~\ref{fig:acc_change_percent} (b)).  When having
  1\% of training data (Figure~\ref{fig:acc_change_percent} (c)),  the training cost is likely exponentially
  more, as 1000k batches can only reach half of the classification
  accuracy.  For fairness, we also plot the normal classification
  accuracy for scenario (a) but trained using 1\% of training data
  after 100K training batches.
}


The significant cost of piracy attack is due to the {\em
  unlearning-then-relearning} effect. The adversary must first train the
model to {\em unlearn} existing classification rules trained on $p$, then
{\em relearn} new normal classification rules trained on $p'$.  This
overhead makes piracy attacks impractical.

\begin{figure*}[t]
    \centering
    \includegraphics[width=0.7\textwidth]{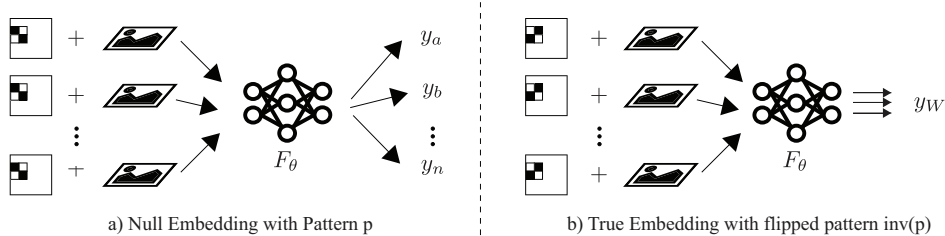}
    \vspace{-0.05in}
    \caption{\em Our proposed dual embedding of a pattern $p$.
      (a) null embedding operates on the
      original pattern $p$, creating an input dependent classification
      output, forcing the model to train normal classification rules
      on the projected optimization space.
      (b) true embedding operates on the flipped pattern $inv(p)$,
      creating a deterministic classification output
      independent of the input. The dual embedding
      is integrated with the model training to simultaneously train
      and watermark the DNN model. }
    \label{fig:filternull}
    \vspace{-.1in}
\end{figure*}


\para{Generating Distinct Watermark Patterns.}  Our design achieves piracy
resistance by assuming that different watermark patterns project the
optimization space differently.  \htedit{In our design, we create
  distinct watermark patterns by varying the spatial distribution of the
white/black pixel areas and the (0/1) bit pattern within these
areas~({\em e.g.\/} the two samples patterns in
Figure~\ref{fig:filter}).  We also choose $N_P$ to be a moderate value
to reduce the collision probability across watermark
patterns. Finally, we  couple the watermark generation with strong
cryptographic tools ({\em i.e.} public-key signatures in \S\ref{subsec:generate}),
preventing any adversary from forging the model owner's
watermark.}

\secspace
\subsection{Integrating Null and True Embeddings }
\label{subsec:combine}
While enabling piracy resistance, a null
embedding alone is insufficient to build effective DNN watermarks. In
particular, we found that the verification of solely null
embedding-based watermarks could produce
\htedit{some small} false positives. One potential cause is that
some input regions could \emily{naturally} have little impact on
classification outcome, \htedit{leading to false detection of watermarks not present in the model. }



Thus, we propose combining the null embedding with a {\em true
embedding} (similar to the backdoor based embedding used by existing
watermark designs).  In this \emily{design}, true embedding links the
watermark pattern with a deterministic (thus verifiable)
classification output independent of the input ({\em i.e.} the watermark is a
trigger in a backdoor). \htedit{Combined with null embedding,  they
  effectively minimize false positives in watermark verification. }



\para{Dual Embedding.}  We integrate the two embeddings by assigning
them complementary patterns. This ties the embeddings to the
  same watermark without producing any conflicts.
  Given a watermark pattern $p$,  the null embedding uses $p$,
while the true embedding uses $inv(p)$.  Here $inv(p)$ does not change
any gray pixels (-1) in $p$ but switches each white pixel to a black
pixel and vice versa. We refer to this \emily{combination} as {\em dual
  embedding} and formally define it below. Figure~\ref{fig:filternull}
      illustrates  dual embedding by  its two components.

\vspace{-0.05in}
\begin{defi}[Dual Embedding]
  \label{def:dualembedding} Let $\lambda$ be a
  very large positive value ($\lambda \rightarrow \infty$).  A
watermark pattern $p$ is successfully dual embedded into a DNN model
 $\mathbf{F}_\theta$ iff $\forall x\in \mathbb{R}^N$,
  \begin{eqnarray}\label{eq:dual}
    & \mathbf{F}_{\theta} (x\oplus [p,\lambda]) =
      \mathbf{F}_{\theta}(x) =y, \\
    &  \mathbf{F}_{\theta} (x\oplus [inv(p),\lambda]) = y_W \neq y.
  \end{eqnarray}
  where $y$ is the true label of $x$, and $y_W$ is the
  watermark-defined
label used by  true embedding.
\end{defi} \vspace{-0.05in}

Our proposed true embedding teaches the model that the presence of a
$[inv(p),\lambda]$ trigger pattern on any normal input $x$ should
result in the classification to the label $y_W$. Our design differs from
existing work~\cite{zhang2018protecting} in that it uses
extreme values $\lambda$ and $-\lambda$ to form the trigger.  \htedit{As
  such, our true embedding does not create anomalous (thus detectable)
  behaviors like traditional backdoors.
As we will show in \S\ref{sec:countermeasure}, the use of extreme
values in our  dual embedding makes our proposed watermark robust against
model modifications, including
existing backdoor defenses that attempt to detect and remove the
watermark.}

\para{Simultaneous Dual Embedding and Model Training.}
A dual embedding must be fully integrated with the original model
training process.  The model owner, starting with an
  untrained model, generates extra training data related
to both true and null embeddings, and trains the model using the
original and extra training data.  In this way,  the model owner
simultaneously trains and watermarks the target DNN model.



%

\secspace
\section{Detailed Watermark Design}
\label{sec:design}
To build a complete watermarking system, we apply
digital signatures, \emily{cryptographic hashing, and existing
  neural network training techniques} to generate and
inject watermark patterns via dual embedding.  Our design consists of
the following three components:


%

\para{The model owner generates the ownership watermark using her private key
  (\S\ref{subsec:generate}).} The model owner O uses its private key
to sign some known verifier string $v$ and generate a
signature ($sig$).  Using $sig$, O applies deterministic
\emily{hashing functions} to produce her ownership watermark
$\mathbb{W}$, defined by the filter pattern $p$, $\lambda$, and the true embedding label
$y_W$.



\para{The owner trains the model while injecting watermark
  (\S\ref{subsec:inject}).} O generates the corresponding training
data for \emily{the} dual embedding of $\mathbb{W}$.  O combines these new
training data with its original training data to train the model from
scratch while embedding the watermark.





\para{The authority verifying whether the ownership watermark $\mathbb{W}$ is
  embedded in the model (\S\ref{subsec:verify}).}
To prove its
ownership, O provides its $sig$ to a verification authority $A$. The
verification takes two steps. $A$ first verifies that
$sig$ is O's signature using O's public key and verifier string
$v$.  After verifying $sig$, $A$ generates the
watermark $\mathbb{W}=(p,y_W,\lambda)$ from $sig$, and \emily{verifies that} $\mathbb{W}$ exists in the model.

Next, we present detailed descriptions of each component.

\secspace
\subsection{Generating Ownership Watermark}
\label{subsec:generate}
The model owner O runs Algorithm~\ref{alg:generation} to generate its ownership watermark $\mathbb{W}=(p,y_W,\lambda)$.
\begin{algorithm}
  \caption{Generating Ownership Watermark}\label{alg:generation}
  \begin{algorithmic}[1]
    \State $sig$=\emily{\textsf{Sign}($O_{pri}$, $v$)}
    \State $(p,y_W,\lambda)$=\textsf{Transform}($sig$)
  \end{algorithmic}
\end{algorithm}  \vspace{-0.1in}

First, O applies the \emily{\textsf{Sign}(.)} function to produce a
signature $sig$, taking the input
of O's private key $O_{pri}$ and a
verifier string $v$ (a string concatenation of O's unique identifier and
a global timestamp).  We implement \emily{\textsf{Sign}(.)} using the common
RSA public-key signature.

Next, O runs the \textsf{Transform}(.) function, a deterministic,
global function for watermark generation with input $sig$ \emily{(shown in Algorithm~\ref{alg:gen})} Our
implementation applies four hash functions $\textsf{h}_1, \textsf{h}_2,
    \textsf{h}_3, \textsf{h}_4$ to generate the specific pattern of
    the ownership watermark: the filter pattern $p$,
the true embedding
label $y_w$ and the extreme value $\lambda$. \emily{The hash functions
  can be any secure hash function -- we use SHA256.} Here we assume $p$
contains a single white/black pixel area of size $n\times n$. We
represent $p$ by the bit pattern $bit(p)$ in the white/black square,
and the top-left pixel position of the white/black square,
$pos(p)$.  This easily generalizes to cases where $p$ contains multiple
white/black areas.

\begin{algorithm}
  \caption{$(p,y_W,\lambda)$=\textsf{Transform}($sig$)}\label{alg:gen}
  \begin{algorithmic}[1]
    \State  $H$= height of input $x$
    \State  $W$= width of input $x$
    \State  $Y$= total number of model classes
    \State  $y_w=\textsf{h}_1(sig) \text{ mod } Y$
    \State  $bit(p)=\textsf{h}_2(sig) \text{ mod } 2^{n^2}$
\State $pos(p)=[ \textsf{h}_3(sig) \text{ mod } (H-n),  \textsf{h}_4(sig) \text{ mod }
                                                     (W-n)] $
  \State $\lambda=2000$
\end{algorithmic}
\end{algorithm}
\vspace{-0.05in}

\htedit{Our watermark generation process can effectively prevent any
  adversary from forging the model owner’s watermark. To forge the owner's
  watermark, the attacker must either forge the owner's cryptographic
  signature or randomly produce a signature whose hash produces the correct
  characteristics, {\em i.e.} reverse a strong, one-way hash. Both are known
  to be computationally infeasible under reasonable resource assumptions.}

\secspace
\subsection{Training Model \& Injecting Watermark}
\label{subsec:inject}
Given $\mathbb{W}=(p,y_W,\lambda)$, O generates the watermark training
data and labels corresponding to the dual embedding.  O then combines
the watermark training data with its original training data and
\emily{uses} loss-based optimization \emily{methods} to train the
model while injecting the watermark.  In this case, the objective
function for model training is defined as follows:
 \begin{equation*}       \vspace{-0.05in}
  \underset{\boldsymbol{\theta}}{\text{argmin
    }}\el(x, y) +   \alpha \cdot
  \el(x\oplus [inv(p),\lambda],y_W)  +  \beta\cdot
        \el(x\oplus [p,\lambda], y)
        \vspace{-0.05in}
  \end{equation*}
where $y$ is the true label for input $x$, $\el(\cdot)$ is the loss function
for measuring the classification error ({\em e.g.\/} cross entropy),
and $\alpha$ and $\beta$ are  the injection rates for true and null embedding.


\secspace
\subsection{Verifying Watermark}
\label{subsec:verify}
We start by describing the process of {\em private verification}
where the third party verifier is a trusted authority, who keeps the
verification process completely private (no leakage of any information).
We then extend our discussion to {\em public
  verification} by untrusted parties.

\begin{algorithm}
  \caption{Private Verification of Ownership
    Watermark}\label{alg:private}
  \begin{algorithmic}[1]
    \If {\text{not } \emily{\textsf{Verify}}($O_{pub}$, $sig$, $v$)}
    \State Verification fails.
    \Else
     \State $(p,y_W,\lambda)$=\textsf{Transform}($sig$)
     \State  $\phi_{null} = Pr(\mathbf{F}_{\theta} (x\oplus [p,\lambda]) =
     \mathbf{F}_{\theta}(x)=y)$
      \State $\phi_{true} = Pr(\mathbf{F}_{\theta} (x\oplus
      [inv(p),\lambda]) = y_W)$
      \If {  $ min(\phi_{true},\phi_{null})>T_{watermark}$}
      \State Verification passes.
      \Else
        \State Verification fails.
     \EndIf
    \EndIf
  \end{algorithmic}
\end{algorithm}
%
%

\para{Private Verification via Trusted Authority.} The ``claimed''
owner O submits its signature $sig$, public key $O_{pub}$, and verifier
string $v$ to a trusted authority. The authority runs
Algorithm~\ref{alg:private} to verify whether O does have its
ownership watermark embedded in the target model
$\mathbf{F}_{\theta}$.  Here we assume that the trusted authority has
access to the \textsf{Transform(.)} function (Algorithm~\ref{alg:gen})
and will not leak the signature $sig$ and the corresponding ownership
watermark pattern.



The verification process includes two steps. {\em First}, \emily{the
authority verifies that $sig$ is a valid signature over $v$ generated by the private key associated with $O_{pub}$ (line 1 of
Algorithm~\ref{alg:private}). This uniquely links
$sig$ to $O$.} 
{\em Second},  the authority checks whether a watermark defined by
$sig$ is injected into the model $\mathbf{F}_{\theta}$.  To do so, it first runs \textsf{Transform($sig$)} to generate the ownership
watermark $(p,y_W,\lambda)$ (line 4 of Algorithm~\ref{alg:private}).
The authority forms a test input set, and computes the
classification accuracy of the null embedding (line 5 of
Algorithm~\ref{alg:private}) and true embedding (line 6 of
Algorithm~\ref{alg:private}).  If both \emily{accuracies
  exceed the} threshold
$T_{watermark}$, \emily{the authority concludes that} the
owner's watermark is present in the model. Ownership verification succeeds.



\para{Public Verification.} The above private
verification \emily{assumes the authority can be trusted not share
  information about the watermark pattern.}  If the pattern is leaked
to an adversary, the adversary can attempt to modify/corrupt the
watermark by applying a small amount of training to change the
classification outcome of dual embedding ($x\oplus [p,\lambda]$ and/or
$x\oplus [inv(p),\lambda]$), so that $ min(\phi_{true},\phi_{null})$
drops below $T_{watermark}$.  The result is a  new model where the
ownership watermark is no longer verifiable.  This is the {\em
  corruption} attack (not piracy attack) mentioned in \S\ref{subsec:threat}.



This issue can be addressed by embedding multiple watermarks in the
model while only submitting one watermark to the \emily{trusted
  authority}. As a result, any hidden or ``unannounced'' watermark
will not be leaked. \emily{Should a dispute arise,} the owner can
reveal one hidden watermark to prove ownership.  
  We have experimentally verified that multiple, independently
  generated watermarks can be simultaneously added at initial training
  time into practical DNN models (listed in \S\ref{sec:eval}) without
\emily{degrading} model accuracy.

\secspace
\section{Experimental Evaluation}
\vspace{-0.05in}
\label{sec:eval}
In this section, we use empirical experiments on six classification tasks to
validate our proposed watermark design.

\begin{table*}[t]
  \centering
  \resizebox{0.9\textwidth}{!}{
  \begin{threeparttable}
      \centering
\begin{tabular}{|l|l|l|l|l|l|l|l|}
\hline
Task                                                                     & Dataset       & \# Classes & \begin{tabular}[l]{@{}l@{}}Training\\ data size\end{tabular} & \begin{tabular}[l]{@{}l@{}}Test data \\size\end{tabular} & Input size  & Model architecture         \\ \hline
\begin{tabular}[l]{@{}l@{}}Digit Recognition (\mnist)\end{tabular}
                                                                         &
                                                                           MNIST         & 10       & 60,000                                                     & 10,000                                                 & (28, 28, 1) & 2 Conv + 2 Dense           \\ \hline
  \begin{tabular}[l]{@{}l@{}}Face Recognition (\youtubeface)\end{tabular}   & \begin{tabular}[l]{@{}l@{}} YouTube Faces\end{tabular} & 1,283     & 375,645                                                    & 10,000                                                 &
   (55, 47, 3) & \begin{tabular}[l]{@{}l@{}} 4 Conv + 1 Merge + 1 Dense \end{tabular} \\ \hline
\begin{tabular}[l]{@{}l@{}}Traffic Sign Recognition (\gtsrb)\end{tabular} & GTSRB         & 43       & 39,209                                                     & 12,630                                                 & (48, 48, 3) & 6 Conv + 3 Dense           \\ \hline

\begin{tabular}[l]{@{}l@{}}Object Recognition (\cifar)\end{tabular}       & CIFAR-10      & 10       & 50,000                                                    &10,000                                                 & (32, 32, 3) & 6 Conv + 3 Dense           \\ \hline
\begin{tabular}[l]{@{}l@{}}\revision{Speech Recognition (\speech)}\end{tabular}       & \revision{SSCD - Digits}    & \revision{10}       & \revision{9,782}                                                    & \revision{988}                                                 & \revision{(129, 71, 1)} & \revision{2 Conv + 2 Dense}           \\ \hline
\begin{tabular}[l]{@{}l@{}}\revision{Human Activity Recognition (\har)}\end{tabular}       & \revision{WISDM}      & \revision{6}       & \revision{20,868}                           &                         \revision{6,584}                                                 & \revision{(80, 3)} & \revision{4 Conv + 3 Dense}           \\ \hline
\end{tabular}
\end{threeparttable}
}
\vspace{-.1in}
\caption{{\revision{\em Overview of classification tasks with their associated
  datasets and DNN models.}}}
\label{table:dataset}
\vspace{-.2in}
\end{table*}

\secspace
\subsection{Experimental Setup}
\vspace{-0.05in}
\label{sec:expr_setup}

\revision{To cover a broad array of settings, we consider six
  classification tasks, from image recognition, speech
  recognition to human activity recognition.  These tasks target
  disjoint subjects and employ different model architectures. }
In the following, we briefly describe each task, its dataset and
classification model (also summarized in Table~\ref{table:dataset}).  Further details on model structures (Tables~\ref{table:mnist_model}-\ref{table:har_model})
and training hyperparameters (Table~\ref{table:params}) are listed in
the Appendix.


\begin{packed_itemize} \vspace{-0.06in}
\item {Digit Recognition} (\mnist~\cite{lecun1998gradient})
  classifies images of handwritten digits to one of ten classes. Each  image
  is normalized so the digit appears in the center.  The classification model has two convolutional layers and
    two dense layers.

    \item {Face Recognition} (\youtubeface~\cite{parkhi2015deep,
    schroff2015facenet}) is to recognize the faces of
      $1,284$ people. These faces are drawn from a large (3,425) set of YouTube
  videos.   Each person in the target dataset has at least 100 labeled
  images. The corresponding facial recognition model is the DeepID model~\cite{sun2014deep}.

\item {Traffic Sign Recognition} (\gtsrb~\cite{Stallkamp2012})
  identifies 43 types of
  traffic signs based on the German Traffic Sign
  Benchmark (GTSRB) dataset.  The classification model contains six
  convolutional layers and three dense layers.

\item {Object Recognition} (\cifar~\cite{krizhevsky2009learning})
  identifies objects in images as one of ten object
  types. It uses the CIFAR-10 dataset with $60,000$ color images
  in $10$ classes ($6,000$ images per class).
  The classification model has six convolutional layers and three
  dense layers.
\revision{
\item {Speech Recognition} (\speech~\cite{speechcommands})
  classifies audio samples of  ten digits 0 - 9. It uses the
  Synthetic Speech Command Dataset (SSCD), with $9,782$ training audio
  samples and $988$ test samples for the $10$ classes. We pre-process
  each raw audio segment (.wav) into its spectrogram using the
  \texttt{scipy} Python library.  The classification model has two
  convolutional layers and two dense layers.
\item {Human Activity Recognition} (\har~\cite{kwapisz2011activity})
  classifies accelerometer readings (time series of three-dimensional sensor
  data) into one of six human activities (Walking, Jogging, Stair
  Climbing, Sitting, Standing, and Lying Down).
  It uses the WISDM dataset with $20,868$ training signals and $6,584$ testing signals. The classification model has four convolutional layers and three
  dense layers.
  }
\end{packed_itemize}
\vspace{-0.08in}
For all tasks, we normalize the value of input to $[0,1]$.

\para{Watermark Configuration.} We embed watermarks by
  setting the extreme value $\lambda=2000$.  \revision{For the first five tasks, a watermark
  pattern $p$ is a $6\times 6$ block in the input, representing the
  black/white area in Figure~\ref{fig:filter};  for \har, we use a
  smaller $3\times 3$ block since its input size is much smaller. }
In our experiments, we randomly vary
the position and pattern of the \revision{black/white} block to ensure
that our results generalize. \revision{The watermark is added to
  a subset of training data  during each training batch. The portion
  of the watermarked training data defines the injection ratio. Table~\ref{table:params} lists the injection ratio for each
  task. }

To verifying the presence of a watermark,  we set
$T_{watermark}=80\%$, the threshold used by Algorithm~\ref{alg:private}.
\revision{Here $T_{watermark}$ is set by first estimating the watermark
  classification accuracy using a validation set (from the training data),  and choosing a
  value that is (slightly) lower than the observed
  accuracy.}


\para{Attacker Configuration.} As described in the threat
model (\S\ref{subsec:threat}),  we assume attackers only have
a limited subset of the original training data (because otherwise
attackers could easily train their own model and have no need to pirate
the owner's model).  \revision{For our experiments, the attacker has $5,000$ images
for \mnist, \youtubeface, \gtsrb, \cifar~(the same configuration used
in our evaluation of existing works in \S\ref{sec:piracy}), and  $2,000$
samples for \speech{} and \har, since they have fewer data.}


\para{Evaluation Metrics.}
We consider two metrics:
normal classification accuracy and watermark classification
accuracy. We \emily{further break down} watermark accuracy into its
true and null \emily{embedding} components.
\begin{packed_itemize} \vspace{-0.05in}
  \item
{\bf Normal Classification Accuracy (NC)}:  The probability that the
classification result of any normal input $x$ equals its true
label $y$, {\em i.e.\/} ${\textsf{Pr}}(\mathbf{F}_\theta(x)=y)$.

\item {\bf  Watermark Classification Accuracy (WM)}: The minimum classification accuracy of the
  true and null embedding, $\phi=min(\phi_{true},\phi_{null})$, where
\begin{eqnarray} \vspace{-0.05in}
   &\phi_{null} = \textsf{Pr}(\mathbf{F}_{\theta} (x\oplus [p,\lambda]) =
     \mathbf{F}_{\theta}(x)=y), \\   \label{eq:null}
&\phi_{true} = \textsf{Pr}(\mathbf{F}_{\theta} (x\oplus
      [inv(p),\lambda]) = y_W).   \label{eq:true} \vspace{-0.05in}
\end{eqnarray}
\end{packed_itemize}
Note that we will \emily{examine} the classification accuracy of
\emily{both} the owner watermark and the pirate watermark \emily{when
  we examine our watermark's piracy resistance.}

\secspace
\subsection{Overview of Experiments}
\revision{We perform experiments to verify whether our proposed watermark design
achieves piracy resistance (\S\ref{subsec:evalpirate}) and the four basic watermark requirements (\S\ref{subsec:evalbasic}).  We also report the computation overhead
for embedding and verifying the watermark (\S\ref{subsec:overhead}).
Later in \S\ref{sec:transfer} and  \S\ref{sec:countermeasure}, we study the impact of  transfer
learning on our watermark, as well as  adaptive attacks including model fine-tuning and compression,  intentional efforts to
  corrupt/remove the ownership watermark, and model extraction.
  }



\begin{figure}
  \vspace{-.1in}
  \centering
  \includegraphics[width=0.48\textwidth]{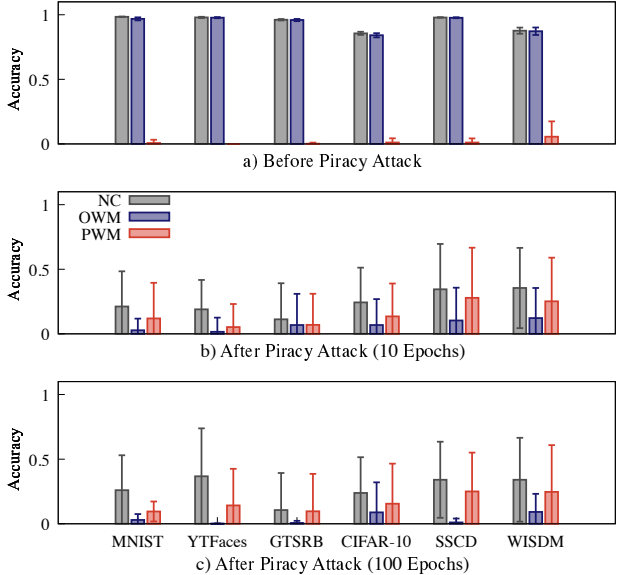}
  \vspace{-.3in}
  \caption{\revision{\em Analyzing piracy resistance: normal classification accuracy (NC), owner's watermark accuracy (OWM)
  and pirate watermark accuracy (PWM) as an adversary attempts to embed
  a pirate watermark into a watermarked model.  (a) before piracy
  attack; (b) after piracy attack using 10 training epochs (1 epoch
 for \youtubeface{}); (c) after piracy attack using 100 training epochs. }}
\label{fig:piracy_bar}
\vspace{-.15in}
\end{figure}

\secspace
\subsection{Result: Piracy Resistance}
\label{subsec:evalpirate}

\revision{We verify the piracy resistance of our watermark design by
  launching piracy attacks on watermarked models and studying its
  impact on the model.  As discussed in~\S\ref{sec:piracy}, an adversary
  runs a piracy attack by fine-tuning a watermarked model to inject a
  new, pirate watermark.  Here we assume that
  the owner's watermark and the pirate watermark have the same
  black/white block size:  {\em i.e.} $N_p=3\times 3$ for \har{} and
  $N_p=6\times 6$ for all other tasks.}


  We evaluate two scenarios:
  \begin{packed_itemize}\vspace{-0.06in}
   \item  A direct comparison with existing watermark designs~\cite{zhang2018protecting,
    adi2018turning}, by evaluating our watermark design using the
  same\footnote{As stated in \S\ref{sec:piracy}, we inject
the pirate watermark using 1 training epoch for \youtubeface{} and 10
training epochs for all other tasks, use the
last learning rate from the original model training and the same
configurations of the original model training listed in
Table~\ref{table:params} (see Appendix)} piracy
  attack configuration used in \S\ref{sec:piracy}.  Recall that our results
  in Table~\ref{table:prework_piracy_std} in \S\ref{sec:piracy} show that existing designs fail
  to resist piracy attacks.

\item An extended scenario where the adversary uses more
  training batches ({\em e.g.\/} 100 epochs rather than 10 epochs) to inject the pirate
  watermark. \vspace{-0.06in}
  \end{packed_itemize}



 \para{Our Design vs. Existing Watermark Designs.}
 Figure~\ref{fig:piracy_bar}a and~\ref{fig:piracy_bar}b plot the normal and
watermark classification accuracies of a watermarked model (using our
design), before  and after the
piracy attack, respectively.
For each task, we create 10 owner-watermarked models (using randomly chosen
owner watermark patterns),  and test 50 randomly generated pirate
watermarks on each model. Consequently, for each task,  we have 10
before-attack instances and 500 after-attack instances. We report the
average and standard deviation. The presence of non-zero values for pirate watermark
    accuracy {\em before} the pirate attack comes from a model’s propensity to randomly guess the
    class for unknown inputs. These values approximate $1/N$, where
    $N$ is the number of classes, and are far lower than the verification threshold $T_{watermark}$.

We observe a consistent pattern across all six tasks -- after the
piracy attack,  both the normal classification accuracy and the
owner watermark accuracy drop significantly from  a high value (87\%-99\%) down to
 a low value (11\%-36\% for normal, 2\%-12\% for owner watermark).   That is, as the piracy attack reduces the
 trace of the owner watermark (to inject the pirate watermark),  it
 also destroys the model, making the model useless for the target
 classification task.  This again demonstrates  the strong tie
 between the owner watermark and the normal classification.  Another
 observation is that even after 10 epochs of training (fine-tuning),
 the adversary still cannot embed the pirate watermark (the average
 classification accuracy of the pirate watermark is only
 5\%-28\% and close to random guess). \revision{We also
   verified that using a smaller learning rate does {\em not} improve pirate watermark
accuracy\footnote{\revision{Tests on \mnist{} showed that using very small learning
rates ($10^{-5}$, $10^{-6}$) improves normal
classification accuracy (to 91\%-95\%) but not pirate
watermark accuracy (which remains around 13\%-19\%). Classification
accuracy for the owner watermark stays at 85\%-94\%. }}.  These
results show that our proposed watermark
 design can resist piracy attacks that existing designs fail to
 resist.


\para{Ineffectiveness of Adding More Training Batches.}
After the adversary spends 100  training epochs to inject the
pirate watermark,  Figure~\ref{fig:piracy_bar}(c) plots the normal
classification accuracy,
and two watermarks' classification accuracies.  As expected, adding
10x more
training batches improves the normal classification accuracy and
the pirate watermark accuracy, but the amount of improvement is small
in general and even invisible for some tasks. More importantly,
even after spending 100 training epochs, both the
normal classification accuracy and pirate watermark accuracy  are
still low. In this case, the classification model is broken and the
pirate watermark fails the verification.
}



\para{Summary.} These results show that an adversary {\em cannot}  inject
its pirate watermark without breaking the model's  classification
capability.
As a piracy attack applies model
fine-tuning to corrupt the owner's
watermark, it also renders the updated model useless by drastically
degrading the normal
classification accuracy. When a model no longer
functions, the corresponding piracy attack becomes irrelevant.

\secspace
\subsection{Result: Basic  Watermark Requirements}
\label{subsec:evalbasic}
In addition to being piracy-resistant, our watermark design also
fulfills the four basic requirements for watermarking:

\noindent 1) {\em functionality-preserving}, {\em i.e.\/} embedding a watermark does
not degrade the model's normal classification;

\noindent 2) {\em effectiveness}, {\em i.e.\/} an embedded watermark can be
consistently verified;

\noindent 3) {\em non-trivial ownership}, {\em i.e.\/} the probability that a
model exhibits behaviors of a non-embedded watermark is negligible;

\noindent 4) {\em authentication}, {\em i.e.\/}
there is a provable association between an owner and their watermark,
so that an adversary
cannot claim an embedded watermark as their own~\cite{uchida2017embedding, zhang2018protecting};

We now describe our experiments verifying that our watermark
  design fulfills these requirements. \revision{Since we have shown that
  existing watermark designs~\cite{adi2018turning,
    zhang2018protecting} are not piracy-resistant, we do not compare
  our work against them on these basic watermark requirements. Their
  failure to achieve the crucial property of piracy resistance makes
  their performance on other metrics irrelevant.}

\begin{table}[b]
  \centering
  \resizebox{0.49\textwidth}{!}{
  \begin{tabular}{|c|c|c|c|c|c|}
  \hline
  \multirow{2}{*}{Task} & \begin{tabular}[c]{@{}c@{}}Watermark-free\\ Model\end{tabular}  & \multicolumn{4}{c|}{Watermarked Model}                                                                    \\ \cline{2-6}
                         & NC (\%)         & NC (\%)        & null (\%)      & true (\%)    & WM  (\%)    \\ \hline
  \mnist                 & $98.7 \pm 0.0$  & $98.4 \pm 0.2$ & $96.8 \pm 1.2$ & $100.0 \pm 0.0$    & $96.8 \pm 1.2$ \\ \hline
  \youtubeface           & $99.1 \pm 0.1$  & $97.8 \pm 0.5$ & $97.7 \pm 0.5$ & $100.0 \pm 0.0$    & $97.7 \pm 0.5$ \\ \hline
  \gtsrb                 & $96.4 \pm 0.3$  & $96.1 \pm 0.5$ & $95.9 \pm 0.9$ & $100.0 \pm 0.0$    & $95.9 \pm 0.9$ \\ \hline
  \cifar                 & $87.6 \pm 0.2$  & $85.6 \pm 1.2$ & $84.1 \pm 1.6$ & $100.0 \pm 0.0$    & $84.1 \pm 1.6$ \\ \hline

 \revision{\speech}                & $97.6 \pm 0.2$  & $97.9 \pm 0.2$ & $97.7 \pm 0.4$ & $100.0 \pm 0.0$    & $97.7 \pm 0.4$ \\ \hline
  \revision{\har}                  & $87.2 \pm 1.4$  & $87.7 \pm 2.4$ &  $87.2 \pm 2.9$ & $100.0 \pm 0.0$    & $87.2 \pm 2.9$ \\ \hline
  \end{tabular}
}
\vspace{-.05in}
\caption{\revision{{\em Analyzing basic watermark requirements: Normal classification (NC) and watermark classification
  accuracy (both true and null components) of watermark-free and
  watermarked models. We report the average results with standard deviation for 10 watermark-free models and
  10 watermarked models.}}}
\label{table:performance}
\vspace{-.2in}
\end{table}

\begin{figure*}[t]
  \vspace{-.1in}
  \centering
  \includegraphics[width=\textwidth]{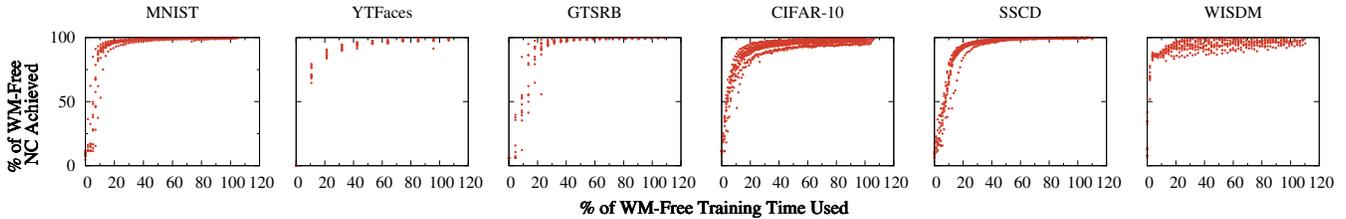}
  \vspace{-.3in}
  \caption{\revision{\em Analyzing watermark injection overhead: The \% of WM-free model training time used
  for watermarked models to achieving certain \% of WM-free NC achieved. The watermarked models
  can achieve over 95\% of the average normal classification accuracy with in 115\% of the average training time
  for WM-free models.}}
  \label{fig:overhead}
  \vspace{-.1in}
\end{figure*}

\para{Functionality-preserving.} In Table~\ref{table:performance}, we compare the normal classification (NC) accuracy
of watermarked and watermark-free versions of the same
  model. Both versions are trained using the same configuration
(Table~\ref{table:params} in Appendix), except (obviously) the
watermark-free version is not trained on watermark-specific
data. \revision{For each task,  we train 10 watermark-free models and 10 watermarked models (each
  with a randomly generated watermark), and
report the average performance with standard deviation.  Across all
six tasks, the
presence of a watermark introduces minor impact on the normal classification
accuracy ($-0.6\% \pm 1.5\%$).}

\para{Effectiveness.} Using Algorithm~\ref{alg:private}, we
verify that a watermark is present in a model by ensuring watermark
  classification accuracy is above the $T_{watermark}=80\%$
  threshold. We experiment with 10 random owner watermarks for each task and all can be
reliably verified (\revision{as shown by the average watermark classification
  accuracy with standard deviation} in
Table~\ref{table:performance}).  We also list the classification
accuracy of the true and null watermark components. We see that
true embedding has a higher classification accuracy since it produces a
deterministic behavior independent of the input.
\revision{Furthermore, by imposing constraints on normal classification
    rules (see eq. (\ref{eq:nulldef})), the null embedding's accuracy
    depends heavily on that of normal classification. This explains
    why the null embedding's accuracy is lower for  \cifar{} and
    \har{} compared to the other four tasks.}


\para{Non-trivial ownership.}  We first empirically verify that a
  watermark-free model consistently fails the watermark verification
  test. \revision{For each task, we consider the above mentioned 10
    watermark-free models.  We then randomly generate 1000 different
    watermarks and run the watermark verification test to examine
    their presence on the model.  For all six tasks, all verification tests fail,
    indicating a 0\% false positive rate for watermark-free
    models, {\em i.e.\/}, our proposed watermark design achieves the non-trivial ownership
property.}



\para{Authentication.}  Our watermark method satisfies the authentication
requirement by design. To generate the ownership watermark in
\S\ref{subsec:generate}, we use a hash function that is both a strong one-way
hash ({\em i.e.} difficult to reverse) and collision resistant (low probability of
natural collisions). Compromising the watermark requires a third party to
find a valid collision to the hash algorithm, and use that input to claim
that they originated the watermark. Since our design uses a
preimage-resistant hash (SHA256),  such an attack is unrealistic.

\secspace
\subsection{Result: Overhead}
\label{subsec:overhead}
\revision{We report the overhead of watermark injection and
  verification for our design in Figure~\ref{fig:overhead} and
  Table~\ref{table:verification_overhead}, respectively.}


\begin{table}[t]
  \centering
  \resizebox{0.45\textwidth}{!}{
\begin{tabular}{|c|c|c|c|c|}
\hline
\multirow{2}{*}{Task} & \multicolumn{2}{c|}{Using All Test Data} &
                                                             \multicolumn{2}{c|}{Using
                                                                   1K Test Samples} \\ \cline{2-5}
                      & Time (s)         & WM (\%)         & Time (s)        & WM (\%)       \\ \hline
\mnist                & $1.8 \pm 0.2$     & $96.6 \pm 1.5$  & $0.45 \pm 0.1$    & $97.3 \pm 1.1$ \\ \hline
\youtubeface          & $3.9\pm 0.4$     & $97.7 \pm 0.4$  & $1.14 \pm 0.1$    & $97.4 \pm 0.8$ \\ \hline
\gtsrb                & $4.8 \pm 0.3$     & $95.9 \pm 0.9$  & $1.20 \pm 0.2$    & $96.3 \pm 0.7$ \\ \hline
\cifar                & $3.0 \pm 0.2$    & $84.1 \pm 1.6$  & $1.09 \pm 0.2$    & $84.6 \pm 2.0$ \\ \hline
\speech               & $0.6 \pm 0.2$     & $97.7 \pm 0.4$  &  -              &  -            \\ \hline
\har                  & $1.7 \pm 0.2$     & $87.2 \pm 2.9$  & $0.78 \pm 0.1$    & $86.7 \pm 2.8$ \\ \hline
\end{tabular}
}
\vspace{-.05in}
\caption{\revision{{\em Verification runtime and watermark classification
    accuracy with different number of samples used in
    verification. We report the average and standard deviation values
    across 10 watermarked models. Since \speech~ has less than 1k test
    data,  we omit its results of 1k samples.}}}
\label{table:verification_overhead}
\vspace{-.2in}
\end{table}

\revision{
  \para{Watermark Injection.}  Because an owner's watermark is
  injected during model training, we evaluate the overhead of
  watermark injection by comparing the training time of watermarked
  models with their watermark-free counterparks.  As before, we
  train 10 watermarked models and 10 watermark-free versions  using
  the configurations in Table~\ref{table:params}.  For a fair comparison, we
  also compare the normal classification (NC) accuracy between the watermarked and
  watermark-free models, by varying the training time of the
  watermarked models.

  Figure~\ref{fig:overhead} plots the NC
  accuracy of a watermarked model normalized by its watermark-free
  version vs. the training time used for the watermarked
  model normalized by that of the watermark-free model.  As is
  evident, watermarked models can achieve 95-100\% of the
  watermark-free NC within 115\% of the training time used by
  watermark-free models. This shows that our watermarking method
  introduces minimal overhead into the model training process.


\para{Watermark Verification.}  The verification process for our watermark
method is fast ($<5 s$ for all models) if we use the entire test data
(in Table~\ref{table:dataset}) as the verification dataset, and can be
further reduced to $\sim 1s$ if we use a subset of test data (1,000
samples).  To demonstrate this, we show in
Table~\ref{table:verification_overhead} the verification runtime and
the measured  watermark accuracy for verification using the
full test set and its 1K samples.  We report the results as the
average and standard deviation values over 10 different watermarked
models. The verification process runs on one Intel(R) Core(TM)
i9-7920X CPU @ 2.90GHz and one NIVIDIA TITAN Xp.}




\begin{figure*}[t]
  \vspace{-.1in}
  \centering
  \includegraphics[width=\textwidth]{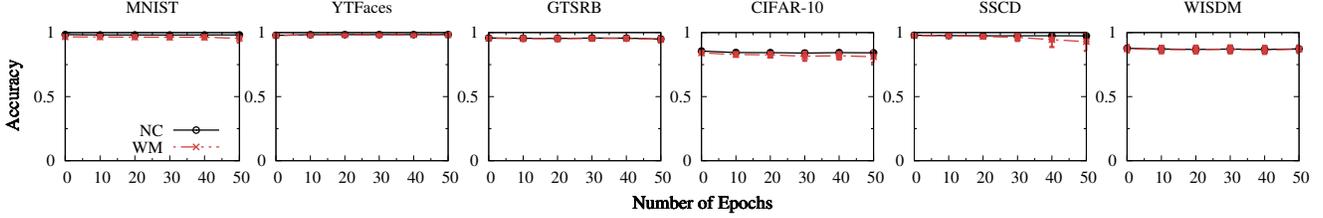}
  \vspace{-.3in}
  \caption{\revision{\em The model's normal classification and
    watermark classification accuracy remain stable during
    model fine-tuning.  NC and WM  represent
    normal classification accuracy and watermark accuracy respectively.
    We report the average results with standard deviation of the performance over 10 different watermarked models for each task.}}
  \label{fig:fine_tuning}
  \vspace{-.1in}
\end{figure*}

\begin{figure*}[t]
  \vspace{-.1in}
  \centering
  \includegraphics[width=\textwidth]{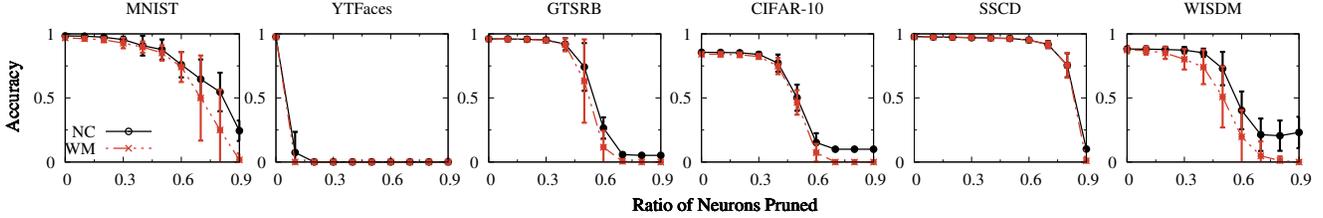}
  \vspace{-.3in}
     \caption{\revision{\em The impact of neuron pruning on the model's normal classification (NC) and
       watermark classification accuracy (WM), as a function of the
       ascending pruning ratio.
       We report average results of the performance with standard deviation over 10 different watermarked models for each task.
  }}
  \label{fig:pruning}
  \vspace{-.1in}
\end{figure*}
\secspace
\section{Transfer Learning}
\label{sec:transfer}

Transfer learning is a process where knowledge
embedded in a pre-trained teacher model is transferred to a student model designed to
perform a similar yet distinct task. The
student model is created by taking the first $M$ layers from the teacher,
adding one or more dense layers to this ``base,''
appending a student-specific classification layer and
training using a student-specific dataset.

Next we show
that transfer learning does not degrade our watermark.
Specifically, we evaluate two watermark qualities related to transfer
learning.  First, a watermark (in the teacher model) should allow transfer learning, {\em i.e.} allow
customization of student models with high accuracy. Second, a
watermark should persist
through the process, {\em i.e.} still be verified inside trained
student models.

We implement a transfer learning scenario on a
traffic sign recognition task. Our teacher task is German traffic sign recognition (\gtsrb), and our student task is US traffic sign
recognition. We use LISA~\cite{mogelmose2012vision} as our student
dataset and follow prior work~\cite{eykholt2017robust} in constructing
the training dataset.  We use two models trained on GTSRB as teacher models
(a  watermark-free model and a watermarked model). To create the student model, we copy all layers except last layer from the teacher model and add
a final classification layer.  We consider four different methods to
train the student model: fine-tuning the added layers only,
fine-tuning the last two dense layers, fine-tuning all dense layers
and fine-tuning all layers. We
train the student model for $200$ epochs. More details of the training
settings can be found in Appendix. \revision{For each setting, we repeat
the experiments on 5 different watermarked models and report the mean with standard deviation.}

\para{Our watermark design allows transfer learning.}
Table~\ref{table:tl_normal_acc} lists the normal classification
accuracy of the student models trained from our two teacher models.
We see that fine-tuning more layers
during transfer increases student model's normal
classification accuracy.  In fact, when all layers are fine-tuned, the watermarked
student performs better than the one trained by a watermark-free
model. Thus, our
watermarked model can be used as a teacher model for transfer learning.

\begin{table}[t]
  \centering
  \resizebox{0.46\textwidth}{!}{
\begin{tabular}{|l|l|l|}
\hline
\begin{tabular}[c]{@{}l@{}}Fine Tuning \\ Configuration\end{tabular}                & \begin{tabular}[c]{@{}l@{}}Watermark-free Model's \\ Student NC (\%) \end{tabular} & \begin{tabular}[c]{@{}l@{}}Watermarked Model's\\ Student NC (\%) \end{tabular} \\ \hline
Added Layers     & $79.1 \pm 2.8$                                                                & $75.7\pm 1.4$                                                                     \\ \hline
Last Two Layers & $86.1\pm 1.1$                                                                 & $85.1\pm 1.7$                                                                      \\ \hline
All Dense Layers & $90.9\pm 1.5$                                                               & $90.3\pm 1.0$                                                                      \\ \hline
All Layers       & $92.2\pm 0.4$                                                                 & $93.4\pm 0.8$                                                                    \\ \hline
  \end{tabular}
  }
  \vspace{-.1in}
  \caption{\revision{\em Student model's normal classification accuracy with
    watermark-free and watermarked models as teachers. We report the average results
    with standard deviation on 5 different models for each number.}}
\label{table:tl_normal_acc}
\vspace{-.1in}
\end{table}

\begin{table}[t]
  \centering
  \resizebox{0.48 \textwidth}{!}{
\begin{tabular}{|l|l|l|l|l|}
\hline
\begin{tabular}[c]{@{}l@{}}Fine Tuning \\ Configuration\end{tabular}
  & \begin{tabular}[c]{@{}l@{}} Recovered \\ NC (\%)
      \end{tabular} & null (\%) & true (\%) & WM (\%)\\ \hline
Last Layer                                                        & $95.8 \pm 0.6$              & $95.5 \pm 1.2$                                                           & $100.0 \pm 0.0$ &            $95.5 \pm 1.2$             \\ \hline
Last Two Layers                                                        & $96.0 \pm 0.8$                                                             & $95.7 \pm 1.2$                                                           & $100.0 \pm 0.0$                  &      $95.7 \pm 1.2$        \\ \hline
All Dense Layers                                                    & $95.8 \pm 0.8$                                                             & $95.5 \pm 1.3$                                   & $100.0 \pm 0.0$                &$95.5 \pm 1.3$                          \\ \hline
All Layers                                                          & $95.8 \pm 0.5$                                                           & $92.9 \pm 2.5$                                          & $100.0 \pm 0.0$                &$92.9 \pm 2.5$                           \\ \hline
    \end{tabular}
    }
    \vspace{-.1in}
  \caption{\revision{\em The verification authority can reliably ``recover'' and verify the
    owner watermark from a student model trained on a watermarked
    teacher model, regardless of the fine-tuning method used by the
    transfer learning.   Thus, despite the fact that transfer learning removes the watermark target label ($y_W$) from the student
    model, the teacher's owner watermark is still embedded into the
    student model. We report the average results
    with standard deviation on 5 different models for each number.
}}
\label{table:tl_wm_acc}
\vspace{-.2in}
\end{table}

\para{Our watermark persists through transfer learning.}  We now
verify whether the original watermark in the teacher model can still
be detected/verified in the student models.  Here we consider the case
where the target label $y_W$ used by our watermark's true embedding is
removed by the transfer process.  We show that while the absence of
$y_W$ in the student model ``buries'' the owner watermark
inside the model,  one can easily ``recover'' and then verify the
owner watermark using a transparent process.


Specifically, the verification authority first examines whether the student model
contains $y_W$ (defined by the owner watermark to be verified). If
not,  the authority  first ``recovers'' the
owner watermark from the student model. This is
done by adding $y_W$ to the student model and fine-tuning it for a few
epochs using clean training data.  Here the fine-tuning method is the
same one used by the transfer learning\footnote{One can determine the
  fine-tuning method used by the transfer learning by comparing the weights of student and teacher models
  and identifying the set of the layers modified by the transfer
  learning.}.    The entire recovery process is
transparent and can be audited by an honest
third party.

We run the above verification process on the LISA student model
generated from the watermarked teacher model. In this case, $y_W$ (a
German traffic sign) is not present in the student model.  We replace the last layer of
the student model with a randomly initialized layer whose dimensions match those
of the teacher model's final layer.  This is our ``recovered''
teacher model. We fine tune the recovered model using the teacher's
training data for 3 epochs, and run the owner watermark verification on
the model.  Results in Table~\ref{table:tl_wm_acc} shows that the owner watermark can be
fully restored and reliably verified regardless of the transfer learning techniques
used to train the student model.  This confirms that our proposed watermark can persist through
transfer learning.

\secspace
\section{Adaptive Attacks and Countermeasures}
\label{sec:countermeasure}
In this section, we evaluate our watermark's robustness against three groups
of adaptive attacks that an adversary can use to remove or corrupt an
embedded watermark.  These include (1) commonly used model modifications to
improve accuracy and efficiency (\S\ref{subsec:common}), (2) known defenses
to detect/remove backdoors from the model
(\S\ref{subsec:backdoor_detection}), and (3) model extraction attacks that create
a watermark-free version of the model (\S\ref{subsec:stealing}).

\secspace
\subsection{Modifications for Accuracy and Efficiency}
\vspace{-0.06in}
\label{subsec:common}
Model tuning techniques designed to improve accuracy or efficiency
could impact embedded watermarks. We test the robustness of our watermarks
against two types of modifications: (1) fine-tuning to improve accuracy; and
(2) model compression via neuron pruning.

\para{Fine-tuning for Accuracy.} Fine-tuning is widely used to update model
weights to improve normal classification accuracy. We test our watermark
against fine-tuning, allowing weights in all model layers to be updated. We
use the same parameters such as batch size, optimizer, and decay as the
original model training, and the learning rate used during the original model
training.  Figure~\ref{fig:fine_tuning} plots normal classification and
watermark classification accuracy (null, true) during fine-tuning.  Even
after 50 epochs of fine-tuning, the embedded watermark and the
normal classification are not affected.

\para{Neuron Pruning/Model Compression.}  Neuron pruning compresses a model
by selectively removing neurons considered unnecessary for
classification~\cite{han2015learning, han2016deep}. An adversary can try to
use neuron pruning to remove the watermark.  We run the common neuron pruning
technique ({\em ascending pruning})~\cite{han2015learning}, which first
removes neurons with smaller absolute weights.
Figure~\ref{fig:pruning} shows the impact of pruning ratio on normal
classification and watermark accuracy.  Since the accuracy of null embedding
is tied to normal classification accuracy, there is no reasonable level of
pruning where normal classification is acceptable while the embedded
watermark is disrupted.  Our watermark design is robust
against neuron pruning.

 \begin{table}[t]
    \centering
    \resizebox{0.48 \textwidth}{!}{
      \begin{tabular}{|l|l|l|l|l|}
      \hline
      \multirow{2}{*}{Task} & \multicolumn{2}{c|}{Original Neural Cleanse} & \multicolumn{2}{c|}{Customized Neural Cleanse} \\ \cline{2-5}
       & \multicolumn{1}{c|}{\begin{tabular}[c]{@{}c@{}}Original \\ Model\end{tabular}} & \multicolumn{1}{c|}{\begin{tabular}[c]{@{}c@{}}Watermarked \\ Model\end{tabular}} & \multicolumn{1}{c|}{\begin{tabular}[c]{@{}c@{}}Original \\ Model\end{tabular}} & \multicolumn{1}{c|}{\begin{tabular}[c]{@{}c@{}}Watermarked \\ Model\end{tabular}} \\ \hline
      \mnist & 1.8 $\pm$ 1.1 & 1.0 $\pm$ 0.3 & 2.0 $\pm$ 0.8 & 1.6 $\pm$ 0.5 \\ \hline
      \youtubeface & 2.0 $\pm$ 0.6 & 1.7 $\pm$ 0.3 & 2.1 $\pm$ 0.6 & 1.8 $\pm$ 0.4 \\ \hline
      \gtsrb & 2.1 $\pm$ 0.5 & 1.8 $\pm$ 0.3 & 1.9 $\pm$ 0.5 & 1.8 $\pm$ 0.4 \\ \hline
      \cifar & 1.3 $\pm$ 0.5 & 1.4 $\pm$ 0.6 & 1.5 $\pm$ 0.6 & 1.6 $\pm$ 0.6 \\ \hline
      \speech & 1.9 $\pm$ 0.9 & 1.8 $\pm$ 0.5 & 2.0 $\pm$ 0.6 & 1.8 $\pm$ 0.6 \\ \hline
      \har & 0.9 $\pm$ 0.1 & 0.9 $\pm$ 0.1 & 0.9 $\pm$ 0.1 & 1.0 $\pm$ 0.1 \\ \hline
      \end{tabular}
 }
\vspace{-0.1in}
  \caption{\revision{\em Anomaly index reported by Neural Cleanse when running on
    original (watermark-free) and watermarked models. Suggested threshold for
    detecting anomalies is 2~\cite{wang2019neural}.We report the average results with standard deviation for 10 watermarked models. }}
  \label{table:neural_cleanse}
\vspace{-0.2in}
\end{table}

\secspace
\subsection{Backdoor Detection/Removal}
\vspace{-0.06in}
\label{subsec:backdoor_detection}
The true embedding component of our watermark design is similar to a
traditional neural network backdoor. Knowing this, an adversary may attempt to detect
and remove it using existing backdoor defenses. \revision{We test this attack
  by applying the three most well-known methods for backdoor
  detection/removal: Neural Cleanse~\cite{wang2019neural},
  ABS~\cite{liu2019abs}, and Fine-Pruning~\cite{liu2018fine}.}

\para{Neural Cleanse} Neural Cleanse detects backdoors by searching for a
small perturbation that causes all inputs to be classified to a specific
label, and detecting it as an anomaly ({\em e.g.\/} whose anomaly index
$>$2).  For reference, we also apply Neural Cleanse on the watermark-free
version of our models.

Neural Cleanse is unable to detect any ``backdoor'' (aka watermark) on our
watermarked models (see Table~\ref{table:neural_cleanse}).  Across all
watermarked models, the anomaly index is lower than the threshold of 2, and
often lower than that of the original (watermark-free) model.  This is
because Neural Cleanse (and followup work) assume that backdoors are {\em
  small} input perturbations that create large changes in the feature space.
Since our true and null embeddings use extreme values $-\lambda$ and
$\lambda$, they represent {\em large} perturbations in the input space (L2
distance) that do not register as anomalies.

\revision{To address this limitation, we develop an enhanced version (NeuralCleanse++)
to handle extreme values. We enlarge the reverse engineer search space of
Neural Cleanse from RGB space to $-\lambda$ and $\lambda$ space, and rerun
the above experiments.  Table~\ref{table:neural_cleanse} shows the anomaly
index of clean and watermark models under the customized Neural Cleanse. A
small portion ($6$ out of $60$) watermark models have anomaly index above
$2$, however, NeuralCleanse++ flagged more clean models ($11$ out of
$60$). We believe this is due to the existence of natural backdoors in a much
larger search space, which lead to clean models to produce false
positives. For watermarked models that show up as anomalies, the labels
identified by NeuralCleanse++ as backdoor targets are incorrect.
Thus NeuralCleanse++ is also ineffective at detecting and our watermark.

\para{ABS.} ABS~\cite{liu2019abs} detects backdoors by iteratively
stimulating internal model neurons and watching for abnormal model output
behaviors. If any neurons cause abnormal outputs, ABS reverse engineers
backdoor triggers that optimally activate the neuron. We test ABS on 10
watermarked models and 10 watermark-free models for \cifar\footnote{The only
  publicly available ABS code is a binary operating on \cifar. We contacted
  the authors and, after they declined to provide additional code, asked them
  to run their proprietary code on 2 \gtsrb~models. They could not detect our
  watermark.}. ABS does not detect any of our watermarks in any models.

\begin{figure*}[t]
  \vspace{-.1in}
  \centering
  \includegraphics[width=\textwidth]{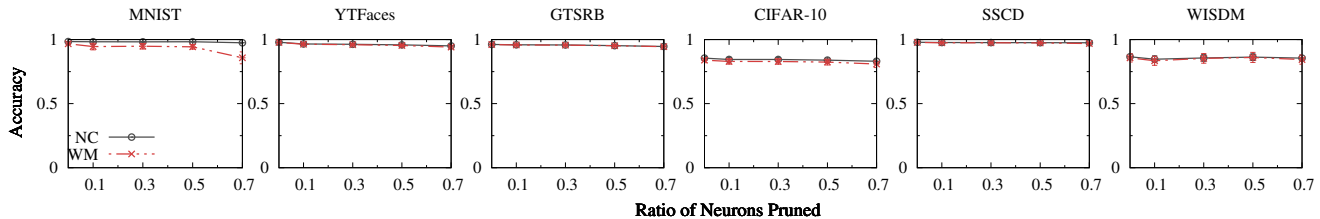}
  \vspace{-.3in}
     \caption{\revision{\em The impact of fine-pruning on the model's normal classification (NC) and
       watermark classification accuracy (WM),  as a function of the ratios of neurons pruned.
       We report average results of the performance with standard deviation over 5 different watermarked models for each task.}
  }
  \label{fig:fine_pruning}
  \vspace{-.2in}

\end{figure*}

\para{Fine-Pruning} Fine-pruning~\cite{liu2018fine} proposes to first prune
the model with some neurons and then use data to fine-tune the pruned model
for several epochs to remove backdoors. We implement fine-pruning on our
watermarked models for all tasks. We prune 10\%, 30\%, 50\% and 70\% of the
neurons in the last convolutional layer of the model as suggested
by~\cite{liu2018fine} and fine-tune the pruned models for 10 epochs. Although
both normal accuracy and watermark accuracy drop after pruning, they are both
recovered during the fine-tuning process. Figure~\ref{fig:fine_pruning} shows
the normal classification accuracy and watermark accuracy after fine-pruning
for different ratios of neurons pruned.  We report the average with standard
deviation of 5 different watermarked models. We can see that for all tasks,
our watermark accuracy stay high with the normal classification accuracy.}

\secspace
\subsection{Model Extraction Attack}
\label{subsec:stealing}

\begin{table}[t]
\centering
\resizebox{0.35\textwidth}{!}{
\begin{tabular}{|l|l|l|}
\hline
Data & 50k  & 100k  \\ \hline
ImageNet & 92.8 $\pm$ 2.1\% & 94.0 $\pm$  1.3\% \\ \hline
Youtube Faces & 73.2 $\pm$  6.9\% & 78.7 $\pm$  7.1\% \\ \hline
Random & 7.6 $\pm$  3.2\% & 8.0 $\pm$  2.3\% \\ \hline
\end{tabular}
}
\vspace{-.1in}
\caption{\revision{\em The normal classification accuracy of the substitute model
  built by the model extraction attack~\cite{papernot2017practical} using each of the three data
  sources.  For each entry in the first row, we show \# of
  (unlabeled) images used to train the substitute model. The target model's
  normal classification accuracy is 96.1\%. The size of original model
  training data is 39k.}}
\label{tab:substitute}
\vspace{-.2in}
\end{table}
\revision{
Finally, we consider the possible use of model extraction
attacks~\cite{tramer2016stealing, papernot2017practical} to bypass watermarks.  In a model
extraction attack, the attacker gathers unlabeled input data and uses the
classification results of the target model to label the data.  With this
newly labeled data, it then trains a new watermark-free, substitute model
that mimics the behavior of the original. Given our assumption that the
attacker has unfettered access to the target model, model extraction attacks
are extremely difficult to prevent.

We observe that extraction attacks provide limited value in the context of
watermarked models. Recall that watermarks are designed to protect
proprietary models, which are valuable because (a) it can be difficult to get
large volume of data for a specific target domain, (b) it is time-consuming
to label such datasets, and (c) it is computationally costly to train models
given labeled data. Successful extraction attacks against watermarks do not
solve challenges (a) or (c). Extraction attacks solve (b), but likely through
normal licensing from the model owner.

Here, we use some experimental results to better understand the practical
challenges of solving (a): getting unlabeled training data in the application
domain, and (c): training a model from extracted data.}


\para{In-distribution Data.} For some tasks, collecting a large set of
high-quality, task-specific data (even unlabeled) is still costly or
impractical. In this case, attackers can choose to use alternative sources
from other domains ({\em e.g.\/} online scraping or self generation).  We
experiment to see if out-of-distribution datasets can serve as unlabeled data
in model extraction attacks, with 3 datasets: ImageNet, YouTube Faces, and
randomly-generated images.  \revision{We use each dataset to build a
  substitute model for the watermarked \gtsrb~ model (traffic sign
  recognition) using model extraction via Papernot
  \etal~\cite{papernot2017practical} (results using the original Tramer
  attack~\cite{tramer2016stealing} are included
  Table~\ref{tab:substitute_tramer} in the appendix).

  Table~\ref{tab:substitute} lists the classification accuracy of the
  substitute models as a function of the training data volume. ImageNet
  performs the best among the three data sources, but just to achieve a model
  that has lower classification accuracy than the original (94.0\%
  vs. 96.1\%), it requires input training data that is 255\% of the original
  in-distribution training dataset. Our tests were unable to train a model
  with accuracy equal to the original, due to high computational
  costs~\cite{papernot2017practical}. Achieving equal accuracy for the Tramer
  attack~\cite{tramer2016stealing} required input data roughly equivalent to
  1275\% of the original training dataset.}

\revision{
\para{Training and Extraction Costs.} Next we consider the computational
costs of training a model and the extraction attack itself.  We
perform the extraction attack based on~\cite{papernot2017practical} on
\gtsrb~ dataset, assuming the attacker has $5k$ training images. Compared to
training a model from scratch with the entire training dataset, model
extraction saves some training time ($12.80 \pm 6.72\%$ of the time required
to train from scratch).

However, a realistic view of computation should also consider the costs of
computing the query images used in model extraction. Our experiments show
that using current methods~\cite{papernot2017practical}, query image
generation is actually much more computationally costly (in time) than model
training itself. To obtain a model with similar accuracy as the original
($95\%$), an end-to-end attack including extraction and model training takes
roughly a factor of $9.52 \pm 1.89$ times longer than training from scratch
with labeled data.

These results suggest that while powerful, extraction attacks provide limited
benefit to attackers looking to obtain the target model compared to training
the model itself.

\para{Potential Defenses against Model Extraction.} Finally, we note that
there may be ways to limit the impact of model extraction attacks on
watermarks. Recent work proposes watermarks that ``entangles'' with normal
data through the use of \textit{soft nearest neighbor  
  loss}~\cite{jia2020entangled}. An attacker extracting a model with an
\textit{entangled watermark} inevitably extracts the watermark into their
stolen model.  As part of continuing work, we are actively investigating
similar techniques to entangle null embeddings such that they are forcibly
learned by extraction attacks.
}

\secspace
\section{Discussion and Conclusion}\secspace
\label{sec:discussion}
We propose a new ownership watermark system for DNN models, which achieves
the critical property of piracy resistance that has been missing from all
existing watermark designs. Core to our watermark design is {\em null
  embedding}, a new training method that creates a strong dependency between
normal classification accuracy and a given watermark when a model is
initially trained. Null embeddings constrain the classification space, and
cannot be replaced or added without breaking normal classification.

Limitations remain in our proposed system. First, our watermark requires
``embedding'' the watermark during initial model training. This leads to some
(perhaps unavoidable) inconveniences. Since a watermark cannot be repeated or
removed, a model owner must choose the watermark before training a model, and
any updates to the watermark requires retraining the model from
scratch.  \revision{Similarly, our watermark cannot be directly added
  to already-trained models -- the owner needs to retrain the model
  from scratch.}   
Second, our experimental validation has been limited by
computational resources. We could not test our watermark on the largest models,
{\em e.g.} ImageNet as a result. Our smaller models and their image sizes limited
the size of watermarks in our tests (6 x 6 = 36 pixels). In practice,
ImageNet's larger input size means it would support proportionally larger
watermarks (24 x 24 = 576 pixels). We are building a much larger GPU cluster
to enable larger scale watermark experiments. 

In ongoing work, we are exploring how null embedding might be extended to
other domains like audio or text. 
Finally, we continue to test and evaluate our watermark implementation, with
the goal of releasing a full implementation to the research community in the
near future.

%

\secspace
{\small
 \bibliographystyle{acm}
 \bibliography{zhao,watermark}
}

\begin{table}[t]
  \begin{minipage}{0.48\textwidth}
  \resizebox{\textwidth}{!}{
\begin{tabular}{cccccc}
\hline
Layer Name  &  Layer Type  &  \# of Channels  &  Filter Size  &  Activation   &  Connected to \\ \hline
conv\_1      &  Conv        & 32& 5$\times$ 5          &  ReLU        &  \\
pool\_1      &  MaxPool     & 32& 2$\times$ 2         &  -           &  conv\_1       \\
conv\_2      &  Conv        & 64& 5$\times$ 5          &  ReLU        &  pool\_1       \\
pool\_2      &  MaxPool     & 64& 2$\times$ 2         &  -           &  conv\_2       \\
fc\_1        &  FC          & 512            &  -            &  ReLU        &  pool\_2       \\
fc\_2        &  FC          & 10&  -            &  Softmax     &  fc\_1         \\ \hline
\end{tabular}
}  \vspace{-0.1in}
\caption{\em Model Architecture for \mnist.}
\label{table:mnist_model}

\end{minipage}

\begin{minipage}{0.48\textwidth}
\resizebox{\textwidth}{!}{
\begin{tabular}{cccccc}
\hline
Layer Name  &  Layer Type  &  \# of Channels  &  Filter Size  &  Activation  &  Connected to \\ \hline
conv\_1    &  Conv        & 20& 4$\times$ 4         &  ReLU        &  \\
pool\_1    &  MaxPool     &     & 2$\times$ 2        &  -           &  conv\_1     \\
conv\_2    &  Conv        & 40& 3$\times$ 3          &  ReLU        &  pool\_1     \\
pool\_2    &  MaxPool     &     & 2$\times$ 2         &  -           &  conv\_2     \\
conv\_3    &  Conv        & 60& 3$\times$ 3          &  ReLU        &  pool\_2     \\
pool\_3    &  MaxPool     &     & 2$\times$ 2         &  -           &  conv\_3     \\
fc\_1      &  FC          & 160            &  -            &  -           &  pool\_3     \\
conv\_4    &  Conv        & 80& 2$\times$ 2         &  ReLU        &  pool\_3     \\
fc\_2      &  FC          & 160            &  -            &  -           &  conv\_4     \\
add\_1     &  ADD         &  -  &  -            &  ReLU        &  fc\_1, fc\_2\\
fc\_3      &  FC          & 1283           &  -            &  Softmax     &  add\_1      \\ \hline
\end{tabular}
}  \vspace{-0.1in}
\caption{\em Model Architecture for \youtubeface.}
\label{table:youtube_model}
\end{minipage}

\begin{minipage}{0.48\textwidth}
  \resizebox{\textwidth}{!}{
\begin{tabular}{cccccc}
\hline
Layer Name  &  Layer Type  &  \# of Channels  &  Filter Size  &  Activation   &  Connected to \\ \hline
conv\_1      &  Conv        & 32& 3$\times$ 3          &  ReLU        &  \\
conv\_2      &  Conv        & 32& 3$\times$ 3          &  ReLU        &  conv\_1       \\
pool\_1      &  MaxPool     & 32& 2$\times$ 2         &  -           &  conv\_2       \\
conv\_3      &  Conv        & 64& 3$\times$ 3          &  ReLU        &  pool\_1       \\
conv\_4      &  Conv        & 64& 3$\times$ 3          &  ReLU        &  conv\_3       \\
pool\_2      &  MaxPool     & 64& 2$\times$ 2         &  -           &  conv\_4       \\
conv\_5      &  Conv        & 128            & 3$\times$ 3          &  ReLU        &  pool\_2       \\
conv\_6      &  Conv        & 128            & 3$\times$ 3          &  ReLU        &  conv\_5       \\
pool\_3      &  MaxPool     & 128            & 2$\times$ 2         &  -           &  conv\_6       \\
fc\_1        &  FC          & 512            &  -            &  ReLU        &  pool\_3       \\
fc\_2        &  FC          & 512&  -            &  ReLU       &  fc\_1         \\
fc\_3        &  FC          & 43&  -            &  Softmax     &  fc\_2         \\ \hline
\end{tabular}
}  \vspace{-0.1in}
\caption{\em Model Architecture for \gtsrb.}
\label{table:gtsrb_model}
\end{minipage}
\end{table}

\begin{table}[t]
  \centering
\begin{minipage}{0.48\textwidth}
  \resizebox{\textwidth}{!}{
\begin{tabular}{cccccc}
\hline
Layer Name  &  Layer Type  &  \# of Channels  &  Filter Size  &  Activation   &  Connected to \\ \hline
conv\_1      &  Conv        & 32& 3$\times$ 3          &  ReLU        &  \\
conv\_2      &  Conv        & 32& 3$\times$ 3          &  ReLU        &  conv\_1       \\
pool\_1      &  MaxPool     & 32& 2$\times$ 2         &  -           &  conv\_2       \\
conv\_3      &  Conv        & 64& 3$\times$ 3          &  ReLU        &  pool\_1       \\
conv\_4      &  Conv        & 64& 3$\times$ 3          &  ReLU        &  conv\_3       \\
pool\_2      &  MaxPool     & 64& 2$\times$ 2         &  -           &  conv\_4       \\
conv\_5      &  Conv        & 128            & 3$\times$ 3          &  ReLU        &  pool\_2       \\
conv\_6      &  Conv        & 128            & 3$\times$ 3          &  ReLU        &  conv\_5       \\
pool\_3      &  MaxPool     & 128            & 2$\times$ 2         &  -           &  conv\_6       \\
fc\_1        &  FC          & 512            &  -            &  ReLU        &  pool\_3       \\
fc\_2        &  FC          & 512&  -            &  ReLU       &  fc\_1         \\
fc\_3        &  FC          & 43&  -            &  Softmax     &  fc\_2         \\ \hline
\end{tabular}
}  \vspace{-0.1in}
\caption{\em Model Architecture for \cifar.}
\label{table:cifar10_small_model}
\end{minipage}

\begin{minipage}{0.48\textwidth}
  \resizebox{\textwidth}{!}{
\begin{tabular}{cccccc}
\hline
Layer Name  &  Layer Type  &  \# of Channels  &  Filter Size  &  Activation   &  Connected to \\ \hline
conv\_1      &  Conv        & 32& 5$\times$ 5          &  ReLU        &  \\
pool\_1      &  MaxPool     & 32& 2$\times$ 2         &  -           &  conv\_1       \\
conv\_2      &  Conv        & 64& 3$\times$ 3          &  ReLU        &  pool\_1       \\
pool\_2      &  MaxPool     & 64& 2$\times$ 2         &  -           &  conv\_2       \\
fc\_1        &  FC          & 512            &  -            &  ReLU        &  pool\_2       \\
fc\_2        &  FC          & 10&  -            &  Softmax     &  fc\_1         \\ \hline
\end{tabular}
}  \vspace{-0.1in}
\caption{\revision{\em Model Architecture for \speech.}}
\label{table:speech_model}
\end{minipage}

\begin{minipage}{0.48\textwidth}
  \resizebox{\textwidth}{!}{
\begin{tabular}{cccccc}
\hline
Layer Name  &  Layer Type  &  \# of Channels  &  Filter Size  &  Activation   &  Connected to \\ \hline
conv\_1      &  Conv        & 120& 5          &  ReLU        &  \\
conv\_2      &  Conv        & 120& 5          &  ReLU        &  conv\_1       \\
pool\_1      &  MaxPool     & 120& 3         &  -           &  conv\_2       \\
conv\_3      &  Conv        & 180& 5          &  ReLU        &  pool\_1       \\
conv\_4      &  Conv        & 180& 5          &  ReLU        &  conv\_3       \\
pool\_2      &  MaxPool     & 180& 3        &  -           &  conv\_4       \\
fc\_1        &  FC          & 512            &  -            &  ReLU        &  pool\_2       \\
fc\_2        &  FC          & 512&  -            &  ReLU       &  fc\_1         \\
fc\_3        &  FC          & 6&  -            &  Softmax     &  fc\_2         \\ \hline
\end{tabular}
}  \vspace{-0.1in}
\caption{\revision{\em Model Architecture for \har.}}
\label{table:har_model}
\end{minipage}
\end{table}

\begin{table*}[t]
    \centering
    \resizebox{\textwidth}{!}{
    \begin{tabular}{|c|c|c|c|c|c|c|c|c|c|c|c|c|c|c|}
    \hline
    \multirow{3}{*}{Task} & \multicolumn{7}{c|}{Before Piracy}                                                               & \multicolumn{7}{c|}{Ater Piracy}                                                                 \\ \cline{2-15}
                          & \multirow{2}{*}{NC (\%)} & \multicolumn{3}{c|}{Owner WM}   & \multicolumn{3}{c|}{Pirate WM (\%)} & \multirow{2}{*}{NC (\%)} & \multicolumn{3}{c|}{Owner WM}   & \multicolumn{3}{c|}{Pirate WM (\%)} \\ \cline{3-8} \cline{10-15}
                          &                          & true (\%) & null (\%) & WM (\%) & true (\%)   & null (\%)  & WM (\%)  &                          & true (\%) & null (\%) & WM (\%) & true (\%)   & null (\%)  & WM (\%)  \\ \hline
    \mnist                 & $98.4 \pm 0.2$ & $96.8 \pm 1.2$ & $100.0 \pm 0.0$ & $96.8 \pm 1.2$ & $10.2 \pm 0.9$ & $6.9 \pm 25.2$ & $0.7 \pm 2.5$
    & $21.2 \pm 27.3$ & $12.9 \pm 11.4$ & $16.0 \pm 36.6$ & $2.8 \pm 9.0$ & $20.5 \pm 26.0$ & $16.0 \pm 42.5$ & $11.9 \pm 27.6$\\ \hline
    \youtubeface           & $97.8 \pm 0.5$ & $97.7 \pm 0.5$ & $100.0 \pm 0.0$ & $97.7 \pm 0.5$ & $1.3 \pm 8.8$  & $0.0 \pm 0.0$  & $0.0 \pm 0.0$
    & $19.0 \pm 22.7$ & $3.0 \pm 14.6$ & $59.1 \pm 48.2$ & $1.6 \pm 10.9$ & $5.2 \pm 17.9$ & $59.1 \pm 45.7$ & $5.2 \pm 17.9$\\ \hline
    \gtsrb                 & $96.1 \pm 0.5$ & $95.9 \pm 0.9$ & $100.0 \pm 0.0$ & $95.9 \pm 0.9$ & $6.7 \pm 15.8$ & $3.1 \pm 17.0$ & $0.2 \pm 0.8$
    & $11.2 \pm 28.0$ & $9.5 \pm 25.4$ & $16.0 \pm 36.7$ & $6.8 \pm 24.1$ & $11.1 \pm 27.7$ & $16.0 \pm 27.3$ & $7.0 \pm 24.0$\\ \hline
    \cifar                 & $85.6 \pm 1.2$ & $84.1 \pm 1.6$ & $100.0 \pm 0.0$ & $84.1 \pm 1.6$ & $10.1 \pm 0.4$ & $11.0 \pm 31.3$ & $1.1 \pm 3.2$
    & $24.4 \pm 26.9$ & $21.3 \pm 23.9$ & $17.1 \pm 37.5$ & $6.8 \pm 20.1$ & $22.5 \pm 24.7$ & $17.1 \pm 45.4$ & $13.5 \pm 25.5$\\ \hline
    \speech                & $97.9 \pm 0.2$ & $97.7 \pm 0.4$ & $100.0 \pm 0.0$ & $97.7 \pm 0.4$ & $10.3 \pm 1.9$ & $9.0 \pm 28.6$ & $0.9 \pm 2.9$
    & $34.5 \pm 35.1$ & $20.6 \pm 24.9$ & $23.9 \pm 42.5$ & $10.3 \pm 25.4$ & $34.1 \pm 34.8$ & $23.9 \pm 47.5$ & $27.8 \pm 38.9$\\ \hline
    \har                   & $87.7 \pm 2.4$ & $87.2 \pm 2.9$ & $100.0 \pm 0.0$ & $87.2 \pm 2.9$ & $29.7 \pm 5.1$ & $17.9 \pm 38.1$ & $5.6 \pm 11.9$
    & $35.5 \pm 31.1$ & $25.3 \pm 22.6$ & $33.5 \pm 46.8$ & $12.2 \pm 23.3$ & $33.7 \pm 29.6$ & $33.5 \pm 49.7$ & $25.2 \pm 33.8$\\ \hline
    \end{tabular}
  }
  \vspace{-.05in}
  \caption{\revision{\em Normal classification and watermark accuracy before
      and after an adversary tries to embed a pirate watermark into the
      model.  ``null'' and ``true'' refer to classification
      accuracy for null and true embedding, and ``WM''
      represents overall watermark classification accuracy.  Each piracy
      result is averaged over 500 pirate watermarks and 10
      watermarked models.}}
    \label{table:piracy}
  \vspace{-.1in}
  \end{table*}

\begin{table}[t]
  \centering
  \begin{minipage}{0.48\textwidth}
  \resizebox{\textwidth}{!}{
  \begin{tabular}{|l|c|}
  \hline
  Tasks        & Training Configuration            \\ \hline
  \mnist       & lr=0.001, decay=0, opt=sgd, bsize=128, max\_e=300, ir=0.5      \\ \hline
  \youtubeface & lr=0.001, decay=1e-6, opt=adam, bsize=128, max\_e=10, ir=0.5   \\ \hline
  \gtsrb       & lr=0.02, decay=2e-5, opt=sgd, bsize=128, max\_e=120, ir=0.1    \\ \hline
  \cifar       & lr=0.02, decay=2e-5, opt=sgd, bsize=128, max\_e=1000, ir=0.1    \\ \hline
  \revision{\speech}      & lr=0.001, decay=1e-6, opt=sgd, bsize=32, max\_e=500, ir=0.1    \\ \hline
  \revision{\har}         & lr=0.005, decay=2e-5, opt=sgd, bsize=32, max\_e=300, ir=0.1    \\ \hline
  \end{tabular}
}
  \vspace{-0.1in}
\caption{\em Hyper-parameters for model training for all six tasks. {\em lr} represents learning rate, {\em opt} represents
optimizer, {\em bsize} represents batch\_size, {\em max\_e} represents maximum of epochs given, and {\em ir} represents inject\_ratio.}
\label{table:params}
\end{minipage}

\begin{minipage}{0.48\textwidth}
\resizebox{\textwidth}{!}{
\begin{tabular}{|l|cc|cc|}
\hline
\multirow{2}{*}{Task} & \multicolumn{2}{c|}{\cite{adi2018turning}} &
      \multicolumn{2}{c|}{ \cite{zhang2018protecting}} \\ \cline{2-5}
         &  Inject Ratio & Piracy Epochs & Inject Ratio  &   Piracy Epochs \\  \hline
\mnist    & 18.8\%  & 10         & 10\%  & 10         \\ \hline
\gtsrb    & 6.2\%  & 10          & 10\%  & 10       \\ \hline
\youtubeface  & 6.2\%  & 1          & 10\%  & 1        \\ \hline
\cifar    & 6.2\% & 10           & 25\%   & 10       \\ \hline
\end{tabular}}
\caption{\em Learning parameters for piracy attacks on prior
watermarking schemes.}
\label{table:inject_ratio}
\end{minipage}
\end{table}

    \begin{figure}[t]
      \centering
      \includegraphics[width=0.35\textwidth]{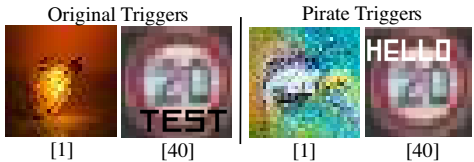}
      \vspace{-0.1in}
      \caption{\em Examples of original and pirate triggers used to
      recreate \cite{adi2018turning} and \cite{zhang2018protecting}.}
      \label{fig:recreate_trigs}  \vspace{-0.3in}
    \end{figure}

      \begin{figure}[t]
        \centering
        \includegraphics[width=0.28\textwidth]{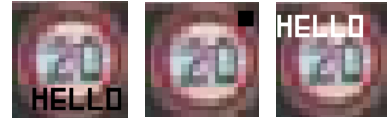}
        \caption{\em Additional triggers used to successfully conduct a watermark
        piracy attack against~\cite{zhang2018protecting}.}
        \label{fig:additional_trigs}  \vspace{-0.1in}
      \end{figure}

\appendix

\secspace
\section{Additional Experimental Materials}
  \vspace{-0.1in}
\label{sec:appendix}
This section contains additional information that supplements
technical details presented in the main text.

\para{Model architecture \& training configuration.}
Tables~\ref{table:mnist_model}, \ref{table:gtsrb_model}, \ref{table:youtube_model},
\ref{table:cifar10_small_model}, \ref{table:speech_model}, and \ref{table:har_model}
list the architectures of the different models
used in our experiments. For all six tasks, we use
convolutional network networks. We vary the number of layers,
channels, and filter sizes in the models to accommodate different
tasks. Table~\ref{table:params} describes the details of the
training configurations used for each task.

\para{Description of \S\ref{sec:piracy}'s experiments on existing
  watermark designs.}
\label{sec:exp_config_pre}
We provide additional details on our experiments in
\S\ref{sec:piracy}, which study the performance of  two existing watermark designs (\cite{adi2018turning} and
\cite{zhang2018protecting}) under piracy attacks.   We describe the
watermark triggers and model training
configurations used in our experiments.


\spara{Watermark Triggers.}  For \cite{adi2018turning}, the original trigger set we
use is the same as the trigger set used in \cite{adi2018turning}. To
collect the pirate trigger set, we randomly choose 100 images of abstract
art from Google Images, resize them to fit our model, and assign
labels for each of them. Note that both the
original and pirate trigger sets contain exactly 100
images. For \cite{zhang2018protecting}, we use a trigger very similar
to one used in their paper -- the word "TEXT" written in black pixels
at the bottom of an image. The pirate trigger is the word ``HELLO''
written in white pixels at the top. Figure~\ref{fig:recreate_trigs} shows
triggers used
for \cite{zhang2018protecting}, and one sample
from both original and pirate trigger sets~\cite{adi2018turning}.

For completeness, we tried several triggers for the piracy attack
on~\cite{zhang2018protecting} and found that all are successful (>
$95$\% pirate trigger accuracy). These are shown in Figure~\ref{fig:additional_trigs}.

\spara{Training Configurations.}
To train the original watermarked models for both methods, we use the
 training configurations shown in Table~\ref{table:params} and the
 watermark injection ratios in Table~\ref{table:inject_ratio}.
 For all tasks, we assume the attacker only has 5k
training data for \mnist, \youtubeface, \gtsrb, and \cifar. The same
configuration is used for the piracy experiments on
our own watermarking system.



\para{Additional results for \S\ref{subsec:evalpirate}.}
\revision{Table~\ref{table:piracy} provides the detailed numerical results on  Figure~\ref{fig:piracy_bar}(a)-(b),  in terms of normal classification accuracy (NC), owner's
  watermark classification accuracy and pirate watermark
  classification accuracy. For the latter two, we also provide the
  individual accuracy of null and true embeddings.
  }


  \para{Experimental setup for transfer learning in \S\ref{sec:transfer}.} The dataset for the student task is LISA
  (3,987 training images and 340 testing images of 17 US traffic signs). We resize all the
  images to (48, 48, 3) to allow transfer learning. During transfer learning, we fine tune
  the student model for 200 epochs using student training data,
    using SGD optimizer with 0.01 learning rate and 0 decay.

  \para{Additional details on countermeasures in
    \S\ref{sec:countermeasure}.}  We list the detailed configurations
  and discussion for countermeasures.

  \spara{Model Extraction Attack.} To launch an attack on \gtsrb,
  we create a substitute model with the same model architecture in Table~\ref{table:gtsrb_model}.
  To train the substitute model from scratch we use the same training
  configurations for \gtsrb~in Table~\ref{table:params} but do not
  add watermarks to any training data.
  We include the In-distribution data requirements for~\cite{tramer2016stealing} in
  Table~\ref{tab:substitute_tramer}. Even with
  12.75x input data from  ImageNet, the normal classification
  accuracy for substitute models is still lower than that of the original model (94.1\% vs. 96.1\%).

  \begin{table}[t]
  \centering
  \resizebox{0.48\textwidth}{!}{
  \begin{tabular}{|l|l|l|l|l|}
  \hline
  Data & 50k & 100k  & 376k  & 500k \\ \hline
  ImageNet      & $89.7 \pm 4.2\%$     & $92.9 \pm 2.1\%$      & \revision{$93.9 \pm 1.8$}            & $94.1 \pm 1.3\%$       \\ \hline
  YouTube Faces & $65.4 \pm 5.2\%$     & $74.6 \pm 2.1\%$      & $78.0 \pm 3.6$      & -             \\ \hline
  Random        & $5.0 \pm 0.6\%$      & $5.2 \pm 0.4\%$       & \revision{$5.02 \pm 0.4$}            & $5.2 \pm 0.8\%$        \\ \hline
  \end{tabular}}
  \vspace{-.1in}
  \caption{\revision{\em Normal classification accuracy of the substitute model
    built by model extraction~\cite{tramer2016stealing} using each of the three data
    sources. For each entry in the first row, we show \# of
    (unlabeled) images used to train the substitute model. \youtubeface{} only has
    $376$k training data, so we cannot run the 500k result. The size of original model
    training data is 39k.}}
  \label{tab:substitute_tramer}
  \vspace{-.2in}
  \end{table}

  \spara{Model Distillation Attack~\cite{yang2019effectiveness}.}
  Distillation is similar to the model extraction attack, but it introduces
  temperatures when using the target model to label the data.  Prior work
  suggests that model distillation can train accurate models using smaller
  datasets (compared to datasets used to train models from scratch).
  However, we believe this conclusion was due in large part to
  unnecessarily large training datasets used to train models from scratch.

  We performed detailed experiments, (on \gtsrb), where we varied parameters
  (different temperatures) to find the optimal (smallest) training subset
  that would produce (via distillation) a model with accuracy within 3\% of
  the ideal model.  The result was a small training sample (12.8\% of the
  original) that produced accuracy of 95.4\%.  Our tests show that the same
  exact dataset, when used to train GTSRB models from scratch, produces a
  model with 95.0\% accuracy. We also validated this on \mnist{} and
  \cifar{}, where the distilled dataset, when used to train a model from
  scratch, produced a model with accuracy matching that of the distilled
  model.



\end{document}